\documentclass{JHEP3}
\usepackage[reqno]{amsmath}
\usepackage{amsfonts,amssymb}
\usepackage{epsfig}
\usepackage{graphicx}
\usepackage{feynmp}

\newcommand{\half}{\frac{1}{2}}
\newcommand{\quarter}{\frac{1}{4}}

\newcommand{\bra}{\langle}
\newcommand{\ket}{\rangle}
\newcommand{\bbra}{\left<\!\left<}
\newcommand{\kket}{\right>\!\right>}
\newcommand{\Tr}{\operatorname{Tr}}

\renewcommand{\Im}{\operatorname{Im}}
\newcommand{\order}{\mathcal{O}}
\newcommand{\sss}{\scriptscriptstyle}

\newcommand{\err}[2]{\raisebox{-0.4ex}
{$\stackrel{\scriptstyle +#1}{\scriptstyle -#2}$}}

\newcommand{\muhat}{\hat{\mu}}
\newcommand{\psibar}{\overline{\psi}}

\newcommand{\gm}{\gamma_\mu}

\newcommand{\smn}{\sigma_{\mu\nu}}
\newcommand{\snl}{\sigma_{\nu\lambda}}

\newcommand{\pslash}{\not\!p}
\newcommand{\qslash}{\not\!q}
\newcommand{\kslash}{\not\!K}
\newcommand{\Kslash}{\not\!K}
\newcommand{\Kt}{\widetilde{K}}
\newcommand{\Ktslash}{\not\!\Kt}
\newcommand{\KdK}{K\!\!\cdot\!\!\Kt}

\newcommand{\Zquark}{Z_2}
\newcommand{\Zgluon}{Z_3}

\newcommand{\z}{{(0)}}
\newcommand{\zz}{Z^{(0)}}
\newcommand{\zmz}{Z^{(0)}_m}
\newcommand{\zmp}{Z^{\sss(+)}_m}

\newcommand{\dmm}{\Delta M^{\sss(-)}}
\newcommand{\csw}{c_{\text{sw}}}
\newcommand{\bm}{b_qm}
\newcommand{\bma}{b_qam}
\newcommand{\bmp}{(1+\bm)}
\newcommand{\bmap}{(1+\bma)}

\newcommand{\ol}{{(1)}}
\newcommand{\gms}{g_{\overline{\text{MS}}}}

\newcommand{\gmomb}{g_{\overline{\text{MOM}}}}
\newcommand{\gmomt}{g_{\widetilde{\text{MOM}}}}
\newcommand{\ams}{\alpha_{\overline{\text{MS}}}}
\newcommand{\amomt}{\alpha_{\widetilde{\text{MOM}}}}
\newcommand{\Lms}{\Lambda_{\overline{\text{MS}}}}
\newcommand{\Lmomb}{\Lambda_{\overline{\text{MOM}}}}
\newcommand{\Lmomt}{\Lambda_{\widetilde{\text{MOM}}}}
\newcommand{\MOMB}{$\overline{\text{MOM}}$}
\newcommand{\MOMT}{$\widetilde{\text{MOM}}$}
\newcommand{\MSB}{$\overline{\text{MS}}$}

\newcommand{\cgluon}{\cite{Leinweber:1998uu}}
\newcommand{\cquark}{\cite{Skullerud:2000un}}
\newcommand{\cquarkII}{\cite{Skullerud:2001aw}}
\newcommand{\cquarks}{\cite{Skullerud:2000un,Skullerud:2001aw}}
\newcommand{\cqqg}{\cite{Skullerud:1997wc}}
\newcommand{\cdos}{\cite{Davydychev:2000rt}}
\newcommand{\cbl}{\cite{Braaten:1981dv}}
\newcommand{\cggg}{\cite{Boucaud:2000ey}}
\newcommand{\cgggs}{\cite{Alles:1997fa,Boucaud:2000ey}}
\newcommand{\cpow}{\cite{Boucaud:2001st}}
\newcommand{\cUKQCD}{\cite{Bowler:1999ae}}
\newcommand{\cLambda}{\cite{Weisz:1995yz}}
\newcommand{\cLambdas}{\cite{Weisz:1995yz,Boucaud:2000ey}}
\newcommand{\csommer}{\cite{Sommer:1994ce,Guagnelli:1998ud}}

\title{Quark--Gluon Vertex from Lattice QCD}

\author{
Jonivar Skullerud\\
Instituut voor Theoretische Fysica, Universiteit van Amsterdam,
Valckenierstraat 65, NL--1018 XE Amsterdam, The Netherlands
}
\author{Ay{\c s}e K{\i}z{\i}lers{\"u}
\\
Special Research Centre for the Subatomic Structure of Matter, 
University of Adelaide, Adelaide SA 5005, Australia}

\abstract{The quark--gluon vertex in Landau gauge is studied in the
quenched approximation using the Sheikholeslami--Wohlert (SW) fermion
action with mean-field improvement coefficients in the action and for
the quark fields.  We see that the form factor that includes the
running coupling is substantially enhanced in the infrared, over and
above the enhancement arising from the infrared suppression of the
quark propagator alone.  We define two different momentum subtraction
renormalisation schemes --- \MOMT\ (asymmetric) and \MOMB\ (symmetric)
--- and determine the running coupling in both schemes.  We find
$\Lms^{N_f=0}=300\err{150}{180}\pm55\pm30$ MeV from the asymmetric scheme.
This is somewhat higher than other determinations of this quantity, but
the uncertainties --- both statistical and systematic --- are large.
In the symmetric scheme, statistical noise prevents us from obtaining
a meaningful estimate for $\Lms$.}

\keywords{Renormalization Regularization and Renormalons, %
Nonperturbative Effects, QCD, Lattice QCD}
\preprint{ITFA 2002-18; ADP-02-72/T511}
\begin{document}

\section{Introduction}
\label{sec:intro}

The quark--gluon vertex plays an important role in many applications
of QCD.
QCD vertex functions may be used to define momentum subtraction (MOM)
renormalisation schemes \cite{Celmaster:1979km,Braaten:1981dv}.
These, 
it is argued, are more `physical' than the minimal subtraction (MS or
\MSB) schemes, since the latter can only be defined in a perturbative
context, while the former are independent of the regularisation method
and give a better guidance to the appropriate renormalisation scale
for a particular problem.
Recently, a complete determination of the quark--gluon vertex to
one-loop order was performed \cdos.  In an asymmetric momentum
subtraction scheme, it has recently been computed to three-loop order
\cite{Chetyrkin:2000dq}, while a numerical computation to two-loop
order has been performed in a symmetric scheme
\cite{Chetyrkin:2000fd}.

The quark--gluon vertex from the lattice may thus yield a direct
determination of the QCD running coupling $\alpha_s$, to complement
other methods for determining this quantity \cLambda.  For a review of
experimental and theoretical determinations of $\alpha_s$, see
\cite{Bethke:2000ai}. 

In addition to reproducing the ultraviolet, perturbative behaviour,
and thus determining the intrinsic QCD scale $\Lms$, the quark--gluon
vertex can also be used to probe the infrared behaviour of the running
coupling.  The hypothesis that the QCD coupling approaches a constant
in the infrared has long been popular on phenomenological grounds, and
has also received some theoretical support from `optimised'
perturbation theory \cite{Mattingly:1994ej}.  On the other hand, a
more recent proposal for reorganising perturbation theory
\cite{VanAcoleyen:2002nd} gives a running coupling that vanishes in the
infrared.  Lattice QCD may in principle assist in resolving this
issue.

The nonperturbative quark--gluon vertex also enters into the
Dyson--Schwinger equations (DSEs)
\cite{Roberts:1994dr,Roberts:2000aa,Alkofer:2000wg}, which are the QCD
field equations.  In particular, in Minkowski space the DSE for the
renormalised quark propagator $S(p)$ is
\begin{equation} 
S^{-1}(p) = Z_2(\pslash - Z_m m)
 + i\frac{4}{3}Z_{1F} g^2\int\frac{d^4k}{(2\pi)^4}  
\gamma_\mu S(k)D^{\mu\nu}(k-p)\Gamma_\nu(p,k-p) \, ,
\label{eq:quark-dse} 
\end{equation} 
where $m$ and $g$ are the renormalised mass and coupling
respectively. 
The unknown quantities here are the nonperturbative gluon propagator
$D_{\mu\nu}(q)$ and the quark--gluon vertex $\Gamma_\mu(p,q)$
[see figure~\ref{fig:vtx-illustrate}].  $Z_2$ and $Z_m$ are the quark
field and mass renormalisation constants respectively.
It will be convenient to also introduce
$\Lambda_\nu(p,q)\equiv-ig\Gamma_\nu(p,q)$. 
\FIGURE{
\includegraphics[width=6cm]{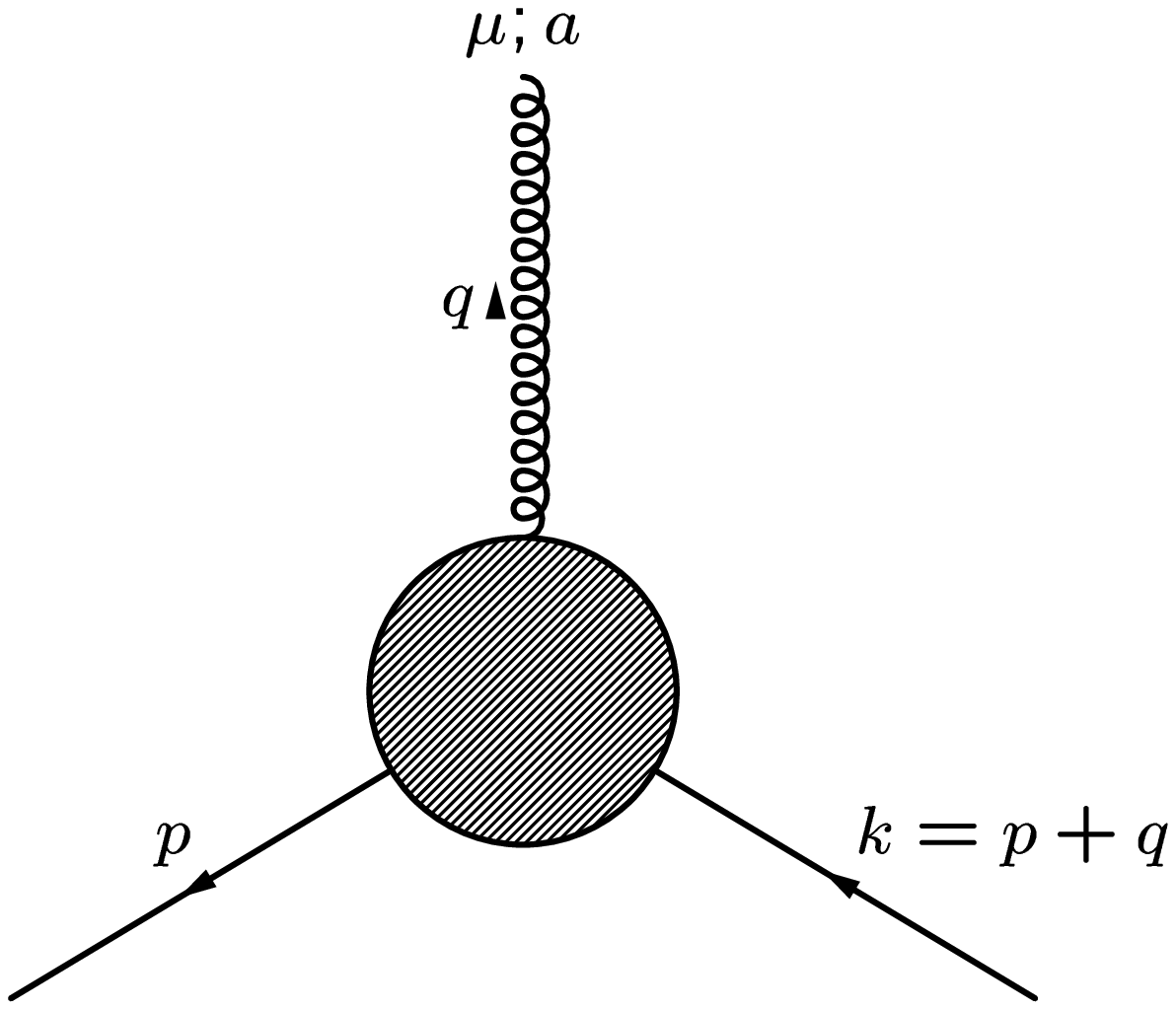}
\caption{The quark--gluon vertex.}
\label{fig:vtx-illustrate}
}

Since these are gauge dependent quantities, it is to be expected that
the confinement picture, and the relative importance of the various
factors, will vary between different gauges.  We will here be working
in the (minimal) Landau gauge, where
over the past few years substantial progress has
been made in our understanding of the gluon propagator from lattice
simulations 
\cite{Leinweber:1998uu,Bonnet:2000kw,Becirevic:1999hj,Langfeld:2001cz}
as well as analytical studies
\cite{vonSmekal:1998is}.

The quark self-energy directly exhibits confinement and dynamical
chiral symmetry breaking, and is an important input for
phenomenological models of hadron physics \cite{Roberts:1994dr}.  It
has also recently been evaluated in Landau gauge on the lattice
\cite{Skullerud:2000un,Skullerud:2001aw,Bowman:2002bm,Bonnet:2002ih}.

However, the quark--gluon vertex remains largely
unknown, and the validity of the usual ans\"atze untested.  In Landau
gauge, there are indications that it must contain non-trivial
structure in the infrared.  The lattice (infrared suppressed) gluon
propagator together with a bare or QED-like vertex fails to yield
solutions to (\ref{eq:quark-dse}) that exhibit an appropriate degree
of dynamical chiral symmetry breaking \cite{Hawes:1998cw}.  But there
are also strong indications that the ghost propagator in Landau gauge
is strongly enhanced, both from lattice simulations
\cite{Suman:1996zg,Cucchieri:1997dx} and analytical studies
\cite{vonSmekal:1998is}.  The ghost self-energy enters
into the quark--gluon vertex through the Slavnov--Taylor identity,
\begin{equation}
q^\mu\Gamma_\mu(p,q) = G(q^2)
 \left[(1-B(q,p+q))S^{-1}(p) - S^{-1}(p+q)(1-B(q,p+q))\right] \, ,
\label{eq:sti}
\end{equation}
where $G(q^2)$ is the ghost renormalisation function and $B^a(q,k) =
t^aB(q,k)$ is the ghost--quark scattering kernel, which is given by
the diagram in figure~\ref{fig:scatt-kernel}.  It appears that modelling
this into the quark DSE does give solutions exhibiting chiral symmetry
breaking and quark confinement \cite{Alkofer:2000mz}.

A nonperturbative determination of the quark--gluon vertex will
therefore give us further insight into the mechanisms of confinement
and chiral symmetry breaking, as well as casting light on the
transition between the perturbative and nonperturbative regimes of
QCD. 
\FIGURE{
\includegraphics[height=5cm]{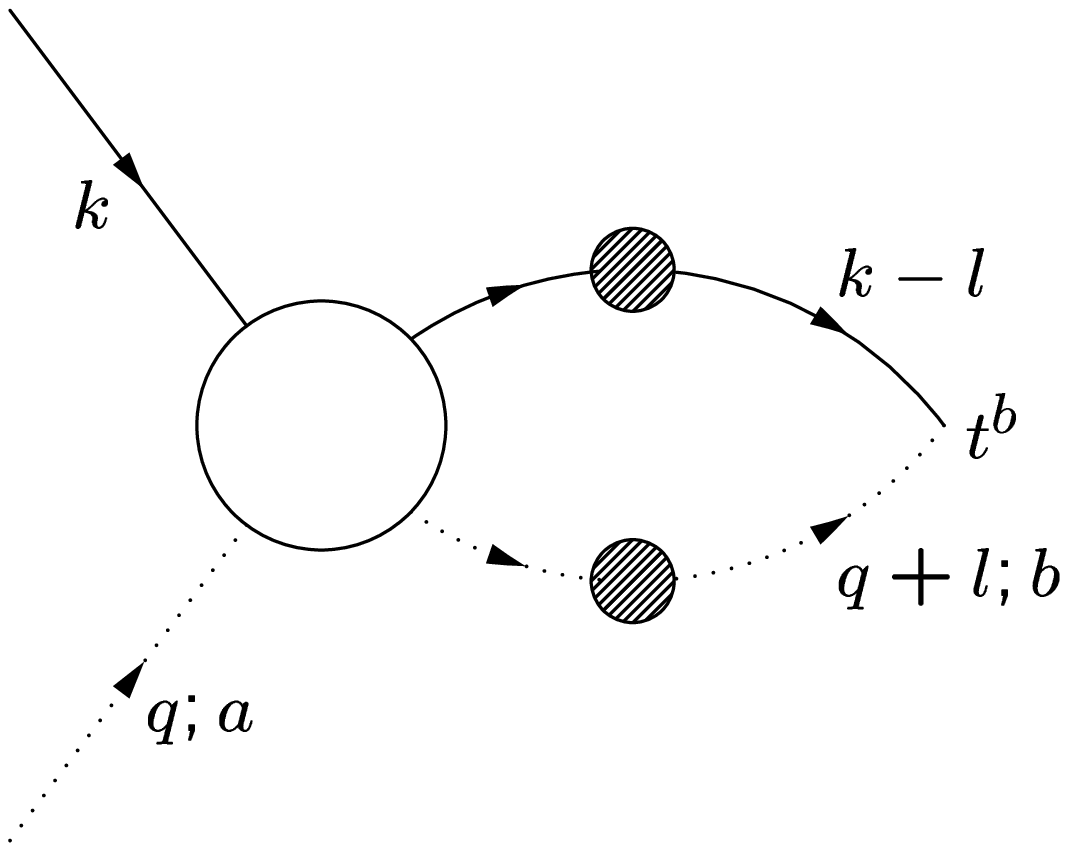}
\caption{The ghost-quark scattering kernel $B^a$.}
\label{fig:scatt-kernel}
}

The starting point for analytical studies of the quark--gluon vertex
is the QED vertex, which gives the `abelian' contribution to the QCD
vertex.  In the abelian case, the Slavnov--Taylor identity implies
that the vertex can be written entirely as a function of the
nonperturbative quark propagator up to a transverse term, as shown by
Ball and Chiu \cite{Ball:1980ay}.  A kinematical basis, along with a
one-loop determination of all the components, is given in
\cite{Ball:1980ay,Kizilersu:1995iz}.

The three-gluon vertex has been the subject of detailed study in
recent years \cgggs.  An important result of these studies is the
discovery of substantial power corrections to the running coupling,
which remain up to scales of 7--10 GeV, and originate from the $\bra
AA\ket$ condensate appearing in the Landau gauge OPE.  It has been
conjectured that this condensate is largely due to instanton effects
\cite{Boucaud:2002nc}.  In the
quark--gluon vertex the situation is in some senses more
straightforward.  Power corrections due to the quark mass appear
already in the one-loop perturbative running coupling, and these are
expected to be substantially enhanced by the chiral condensate.  These
are phenomena that will appear in any gauge.  Any correction due to a
covariant-gauge $\bra AA\ket$ condensate will come in addition to this.

This paper builds on earlier results that were presented in \cqqg.
We will be focusing on the form factor multiplying $\gamma_\mu$, which
contains the running coupling.  We investigate the infrared behaviour
of this form factor, to study the hypothesis of infrared enhancement,
as well as the ultraviolet behaviour, attempting to determine the
perturbative running.

In section~\ref{sec:defs} we define our notation and the quantities
involved.  In section~\ref{sec:renorm} we define two momentum
subtraction schemes based on the quark--gluon vertex, which we call
\MOMT\ and \MOMB, and explain how the running coupling may be
extracted in these schemes.  The parameters of our simulations are
given in section~\ref{sec:comput}.  In section~\ref{sec:zgluon-zquark} we
determine the quark and gluon field renormalisation constants $Z_2$
and $Z_3$.  Our results for the $\lambda_1$ form factor and the \MOMT\
running coupling are given in section~\ref{sec:asymm}, while the results
for the \MOMB\ scheme are given in section~\ref{sec:symm}.  In
section~\ref{sec:match} we translate the MOM scheme results to the \MSB\
scheme.  Finally, in section~\ref{sec:outlook} we summarise our
conclusions and discuss the prospects for further work.  The
appendices contain a discussion of the full tensor decomposition of
the vertex; the lattice tree-level expressions and
our method for removing the dominant (tree-level) lattice artefacts;
and the full one-loop continuum expressions for the form factors we
consider. 

\section{Definitions and principles}
\label{sec:defs}

With the exception of section~\ref{sec:match}, where the perturbative
expressions are given in Minkowski space, we will be working
throughout in euclidean space, with a positive metric, such that
$A^2>0$ for any spacelike vector $A$.  The commutation relations for
the Dirac matrices are the usual ones,
\begin{equation}
\{\gamma_\mu,\gamma_\nu\} = 2\delta_{\mu\nu}\, ; \qquad
\gamma_\mu^\dagger = \gamma_\mu \,.
\end{equation}
The $\smn$-matrices are defined by
\begin{equation}
\smn \equiv\half[\gamma_\mu,\gamma_\nu] \,.
\end{equation}
The generators $t^a$ of the Lie algebra have the conventional
normalisation, $\Tr(t^at^b)=\half\delta^{ab}$.

We can define the configuration space quark--gluon vertex function
(see figure~\ref{fig:vtx-illustrate}) on the lattice as
\begin{equation}
V^a_\mu(x,y,z)^{ij}_{\alpha\beta} = 
\left< \psi^i_\alpha(x)\psibar^j_\beta(z)A^a_\mu(y)\right> 
  = \bbra S^{ij}_{\alpha\beta}(x,z)A^a_\mu(y)\kket \ .
\label{def:vtx-pos}
\end{equation}
Here, $\bra\cdots\ket$ denotes averaging over all fermion and gauge
field configurations, while $\bra\bra\cdots\ket\ket$ denotes averaging
over gauge field configurations only.
Fourier transforming this  and invoking translational invariance gives
us the full (unamputated) momentum space bare vertex function $V_\mu^a(p,q)$:
\begin{multline}
\sum_{x,y,z}
 e^{-i(p\cdot x+q\cdot y-k\cdot z\!)}\!\left<\psi^i_\alpha\!(x)
A^a_\mu\!(y)\psibar^j_\beta\!(z)\right> \\
= \sum_ze^{-i(p-k+q)\cdot z}\sum_{x,y}e^{-i(p\cdot x+q\cdot y)}
\left<\psi^i_\alpha\!(x)A^a_\mu\!(y)
\psibar^j_\beta\!(0)\right> \\
= V\delta(p\!-\!k\!+\!q)
 \bbra S^{ij}_{\alpha\beta}(p;U)A^a_\mu(q)\kket
 \equiv V\delta(p\!-\!k\!+\!q)\;V^a_\mu(p,q)^{ij}_{\alpha\beta} \, ,
\label{def:vtx}
\end{multline}
where $V$ is the lattice volume.
The proper (one-particle irreducible) bare vertex $\Lambda_\mu$ can be obtained
by amputating the external quark and gluon legs from the full vertex
$V^a_\mu$: 
\begin{equation}
\Lambda_\mu^{a,\rm lat}(p,q) = S(p)^{-1}
V^a_\nu(p,q)S(p+q)^{-1}D(q)^{-1}_{\nu\mu}
\, . 
\label{eq:vtx-amputate}
\end{equation}
The only possible dependence this can have on the group coordinates
$a,i,j$ is proportional to the generator $t^a_{ij}$.  We can
therefore consider only $\Lambda_\mu(p,q)$, defined by
\begin{equation}
(\Lambda^a_\mu)^{ij}_{\alpha\beta} =
t^a_{ij}(\Lambda_\mu)_{\alpha\beta}
\equiv -ig_0 t^a_{ij}(\Gamma_\mu)_{\alpha\beta} \, .
\end{equation}
$S(p) = \bbra S(p;U)\kket
 = \bbra\sum_x e^{-ipx}S(x,0;U)\kket$ is the
momentum-space quark propagator, while $D(q)$ is the gluon
propagator, which in the infinite-volume limit takes the form
\begin{equation}
D^{ab}_{\mu\nu}(q) = \delta^{ab}D_{\mu\nu}(q) = 
\delta^{ab}\left(\delta_{\mu\nu}-\frac{q_{\mu}q_{\nu}}{q^2}\right)D(q^2) 
+ \delta^{ab}\xi\frac{q_{\mu}q_{\nu}}{q^2} \frac{1}{q^2} \, .
\label{eq:tensor-infvol}
\end{equation}
In the Landau gauge ($\xi$=0), $D(q^2)$ can be determined for $q\neq0$ by
\begin{equation}
D(q^2) = \frac{1}{3(N^2_C-1)}\sum_{\mu,a} D^{aa}_{\mu\mu}(q) \, .
\label{eq:gluon-infvol}
\end{equation}
As long as $q$ is not too close to zero, this form remains valid also
for a finite volume.  In general, a finite volume will induce an
effective `mass' $m\sim1/L$, which on an asymmetric lattice may also
be direction-dependent --- so the tensor structure
(\ref{eq:tensor-infvol}) must be replaced by \cgluon
\begin{equation}
D^{ab}_{\mu\nu}(q) = 
\delta^{ab}\left(\delta_{\mu\nu}-\frac{h_{\mu\nu}(q)}{f(q^2)}\right)D(q^2) 
+ \delta^{ab}\xi\frac{h'_{\mu\nu}}{g(q^2)}
\equiv \delta^{ab}T_{\mu\nu}(q)D(q) + \delta^{ab}\xi\frac{h'_{\mu\nu}(q)}{g(q^2)} \, ,
\label{eq:tensor-finvol}
\end{equation}
where the functions $f, g, h, h'$ are such that for sufficiently large
$q$, (\ref{eq:tensor-finvol}) approaches the infinite-volume form, but
both $f$ and $g$ remain non-zero as $q\to0$.  The Landau-gauge
expression (\ref{eq:gluon-infvol}) must for the smallest momentum
values --- and in particular for $q=0$ on any volume --- be replaced by
\begin{equation}
D(q^2) = \frac{1}{(N^2_C-1)}\sum_{\mu,a} D^{aa}_{\mu\mu}(q)
 / \sum_\mu T_{\mu\mu}(q)
 \equiv  \frac{1}{T(q)(N^2_C-1)}\sum_{\mu,a} D^{aa}_{\mu\mu}(q)\, .
\label{eq:gluon-finvol}
\end{equation}
In the infinite-volume limit, $T_{\mu\nu}(0)\to\delta_{\mu\nu}$ since
\label{proj0}
the Landau gauge condition places no restriction on the zero-modes
\cite{Cucchieri:1999sz}, so $T(0)\to4$.  In \cgluon\ it was found that an
asymmetric finite volume may induce large distortions to this form,
and in general the components must be determined numerically.
However, $T_{\mu\nu}(0)$ will always remain diagonal.

Since the gluon propagator in Landau gauge for $q\neq0$ becomes
proportional to the transverse projector $P_{\mu\nu}(q) \equiv
\delta_{\mu\nu}-q_{\mu}q_{\nu}/q^2$ or its lattice equivalent,
$D^{-1}$ is undefined and we cannot use eq.~(\ref{eq:vtx-amputate}).
Instead, we rewrite (\ref{eq:vtx-amputate}) as follows,
\begin{equation}
  D_{\mu\nu}(q)\Lambda^a_\nu(p,q)
  = P_{\mu\nu}(q)D(q^2)\Lambda^a_\nu(p,q) = 
S(p)^{-1}V^a_\mu(p,q)S(p+q)^{-1} \, ,
\end{equation}
from which we can obtain the transverse-projected vertex,
\begin{equation}
\Lambda^{a,P}_\mu(p,q) \equiv 
 P_{\mu\nu}(q)\Lambda_\nu^{a,\rm lat}(p,q) = S(p)^{-1}
V^a_\mu(p,q)S(p+q)^{-1}D(q^2)^{-1}
\, . 
\label{eq:vtx-amp-trans}
\end{equation}

The quantities calculated on the lattice are always functions of the
bare (unrenormalised) fields $\psi^0, A_\mu^0$ and the bare coupling
$g_0$.  The relation between renormalised and bare quantities is given
by
\begin{equation}
\psi^0 = \Zquark^{1/2}\psi\,; \qquad \psibar^0 = \Zquark^{1/2}\psibar\,;
\qquad A_\mu^0 = \Zgluon^{1/2}A_\mu\,; \qquad g_0 = Z_g g\, ; \qquad
\xi_0 = Z_3\xi\, ,
\label{eq:renorm-fields}
\end{equation}
where $\Zquark,\Zgluon,Z_g$ are the quark, gluon and vertex (coupling)
renormalisation constants respectively, and are functions of the
regularisation parameter $a$ and the renormalisation scale $\mu$.
From (\ref{eq:renorm-fields}) it follows that
\begin{align}
S^{\text{bare}}(p;a) & = \Zquark(\mu;a)S(p;\mu) \, ;
\label{eq:quark-renorm} \\
D^{\text{bare}}(q^2;a) & = \Zgluon(\mu;a)D(q^2;\mu) \, .
\label{eq:gluon-renorm}
\end{align}
The renormalised vertex is related to the bare vertex according to
\begin{equation}
\Lambda_\mu^{\text{bare}}(p,q;a)
 = Z_{1F}^{-1}(\mu;a)\Lambda_\mu(p,q;\mu) \, .
\end{equation}
Gauge invariance requires that $Z_{1F} = Z_g Z_2 Z_3^{1/2}$, and so we
may write
\begin{equation}
\Lambda_\mu^{\text{bare}}(p,q;a) = Z_g^{-1}(\mu;a)
\Zquark^{-1}(\mu;a)\Zgluon^{-1/2}(\mu;a)\Lambda_\mu(p,q;\mu) \, ,
\label{eq:vtx-renorm}
\end{equation}
meaning that only the quark and gluon fields, along with the running
coupling, are independently renormalised.  For the sake of brevity, we
will from here on no longer explitly label bare quantities as such.
$\Zquark$ and $\Zgluon$ may be determined by imposing momentum
subtraction (MOM) renormalisation conditions on the quark and gluon
propagator respectively, demanding that they take on their tree-level
forms at the renormalisation scale $\mu$:
\begin{align}
S(p;\mu)\bigm|_{p^2=\mu^2} & =
 \frac{1}{i\pslash + m(\mu)} \, ; \label{eq:Zquark-def} \\
D(q^2;\mu)\bigm|_{q^2=\mu^2} & =
 \frac{1}{\mu^2} \, . \label{eq:Zgluon-def} 
\end{align}
The renormalisation of the coupling will be discussed in
section~\ref{sec:renorm}.

The Lorentz structure of the vertex in the continuum consists of 12
independent vectors and can be written in terms of a `Slavnov--Taylor'
(non-transverse) and purely transverse part in terms of vectors
$L_i,T_i$ and scalar functions $\lambda_i,\tau_i$:
\begin{equation} 
\begin{aligned}
\Lambda_\mu(p,q) &= 
 \Lambda_\mu^{(ST)}(p,q) + \Lambda_\mu^{(T)}(p,q) \\
& = -ig\sum_{i=1}^{4}\lambda_i(p^2,q^2,k^2)L_{i,\mu}(p,q) 
    -ig\sum_{i=1}^{8}\tau_i(p^2,q^2,k^2)T_{i,\mu}(p,q) \, .
\end{aligned}
\label{eq:decompose}
\end{equation}
The full expressions for all the vectors $L_i$ and $T_i$ are given in
appendix~\ref{sec:decompose}.  In this paper, we will only study the
part of the vertex proportional to $\gamma_\mu$, which in the specific
kinematics we will be employing is given by the three vectors
\begin{equation}
L_{1,\mu} = \gamma_\mu \,; \qquad
L_{2,\mu} = -(2\pslash+\qslash) (2p+q)_\mu  \,; \qquad
T_{3,\mu} = \qslash q_\mu - q^2\gamma_\mu \,.
\end{equation}
Because of the Slavnov--Taylor identity (\ref{eq:sti}), the scalar
functions $\lambda_i(p^2,q^2,k^2)$ in (\ref{eq:decompose}) may be
expressed in terms of the quark propagator, ghost propagator and
ghost--quark scattering kernel.  In QED for instance, as a result of
the Ward--Takahashi identity, $\lambda_1$ is given uniquely in terms
of the fermion propagator $S^{-1}(p) = i\pslash
A(p^2)+B(p^2)$\footnote{Note that the opposite conventions for the
$B$-function are often used in Minkowski space: our $B$ corresponds to
$-\beta$ in \cite{Davydychev:2000rt,Ball:1980ay}} by
\cite{Ball:1980ay}
\begin{equation}
\lambda_1^{\text{QED}}(p^2,q^2,k^2)
 = \half\left(A(p^2) + A(k^2)\right) \,.\label{eq:bc1}
\end{equation}
In QCD, for $q=0$, the equivalent of this is \cite{Saks:PhD}
\begin{multline}
\lambda_1(p^2,0,p^2) = G(0)\Bigl[A(p^2)\chi_0(p^2,0,p^2)
 + B(p^2)(\chi_1(p^2,0,p^2)+\chi_2(p^2,0,p^2) )\\
 - 2p^2A(p^2)\chi_3(p^2,0,p^2)\Bigr] \, ,
\end{multline}
where $\chi_i$ are the form factors of the ghost-quark scattering
kernel given in \cdos.  At tree level, $\chi_0=1, \chi_{1,2,3}=0$.

As already mentioned, in Landau gauge, for $q^2\neq0$, we can only
determine the transverse part of the vertex from the lattice.  The
transverse projection of $L_1$ is
\begin{equation}
P_{\mu\nu}(q)L_{1,\nu}(p,q) = -\frac{1}{q^2}T_{3,\mu}(p,q)
\label{eq:project} \, .
\end{equation}
This gives rise to the modified form factor $\lambda'_1$,
\begin{equation}
\lambda'_1 = \lambda_1 - q^2\tau_3 \label{def:lambdaprime} \, ,
\end{equation}
which will be useful when studying the transverse-projected vertex.

\section{Definition of the MOM schemes}
\label{sec:renorm}

We impose the momentum subtraction scheme
\begin{equation}
  \lambda_1(\mu) = 1 \, ,
\end{equation}
where `$\lambda_1(\mu)$' stands for $\lambda_1$ evaluated at a specific
kinematic point (e.g., symmetric or zero-momentum), with the momentum
scale $\mu$.  The precise meaning of this will be clear when we
discuss the \MOMT\ and \MOMB\ schemes.  It then follows from
(\ref{eq:vtx-renorm}) that
\begin{equation}
\begin{split}
  g_R(\mu) = Z_g^{-1}(\mu;a)g_0(a)
& = Z_2(\mu;a) Z_3^{1/2}(\mu;a)g_0(a)
\frac{\lambda_1^{\text{bare}}(\mu;a)}{\lambda_1(\mu)} \\
& = Z_2(\mu;a) Z_3^{1/2}(\mu;a) g_0(a)\lambda_1^{\text{bare}}(\mu;a) \, .
\end{split}
\end{equation}
$g_0\lambda_1^{\text{bare}}$ is the quantity we calculate on the lattice.

We will define two different renormalisation schemes, \MOMT\ and
\MOMB.  The `asymmetric' \MOMT\ scheme is defined by setting the gluon
momentum $q^2$ to zero.  This differs from the \MOMT\ scheme defined
in \cbl, and also from the \MOMT$_q$ scheme defined and computed to
three-loop order in \cite{Chetyrkin:2000dq}, where in both cases one
of the quark momenta has been set to zero.  The `symmetric' \MOMB\
scheme is defined by the kinematics $p=-k=-q/2=s$, so
$p^2=k^2=q^2/4=s^2$. 
The fully symmetric scheme where $p^2=q^2=k^2$ is impossible to
implement on a finite lattice where the boundary conditions are
different for fermions and gauge fields (antiperiodic and periodic in
time respectively), which is why we are not considering it here.

In any scheme, the first step towards extracting the running coupling,
which is proportional to $\lambda_1$, is to eliminate those form
factors with a different Dirac structure by tracing the vertex with
$\gamma_\mu$.  With this in mind, we define the functions $H_\mu(p,q)$
as
\begin{align}
H_\mu(p,q) & \equiv -\quarter\Im\Tr \gamma_\mu \Lambda_\mu(p,q)
\notag \\
\begin{split}
 & = g_0\Bigl\{\lambda_1 - (2p+q)_\mu^2\lambda_2 
  + [(p\cdot q)q_\mu-q^2p_\mu](2p+q)_\mu\tau_2 \\
 & \phantom{=g_0\Bigl\{} - (q^2-q_\mu^2)\tau_3
 - [q\cdot(2p+q) - q_\mu(2p+q)_\mu]\tau_6\Bigr\} \, ,
\label{eq:Kmu-expand}
\end{split}
\end{align}
where no sum over $\mu$ is implied.

In the \MOMT\ scheme, all terms in (\ref{eq:Kmu-expand}) proportional
to $q$ or $q_\mu$ disappear, and we are left with
\begin{equation}
H_\mu(p,q=0) = g_0\Bigl(\lambda_1(p^2,0,p^2) - 4p_\mu^2\lambda_2(p^2,0,p^2)\Bigr) \, .
\end{equation}
We can then eliminate $\lambda_2$ by imposing an appropriate
kinematics: $p_\mu=0, p_\nu\neq 0$ for $\mu\neq\nu$.  This defines
$\lambda_1^{\widetilde{\text{MOM}}}(\mu)\equiv\lambda_1(\mu^2,0,\mu^2)$.

The \MOMT\ renormalised coupling is then defined as
\begin{equation}
\gmomt(\mu) = \Zquark(\mu)\Zgluon^{1/2}(\mu)
g_0\lambda_1(\mu^2,0,\mu^2) \, .
\label{eq:gr-asym}
\end{equation}

In the \MOMB\ scheme, the transverse-projected vertex gives us
\begin{equation}
\begin{aligned}
H^T_\mu(p,-2p) &= g_0(4p^2-4p_\mu^2)\Bigl(\frac{1}{4p^2}\lambda_1(p^2,4p^2,p^2) 
 - \tau_3(p^2,4p^2,p^2)\Bigr) \\
 &\equiv g_0\Bigl(1-\frac{p_\mu^2}{p^2}\Bigr)\lambda'_1(p^2,4p^2,p^2)  \, .
\end{aligned}
\end{equation}
$\lambda'_1$ can then easily be extracted by
\begin{equation}
\lambda'_1(p^2,4p^2,p^2)
 = \frac{1}{3}\sum_\mu H_\mu(p,-2p) \equiv \frac{1}{3}h_1(p^2) \, .
\label{eq:sym-k1}
\end{equation}
Thus, we define the \MOMB\ running coupling as
\begin{equation}
\gmomb(\mu) = 
\Zquark(\mu)\Zgluon^{1/2}(\mu)g_0\lambda'_1(\mu^2,4\mu^2,\mu^2) \, .
\label{eq:gr-sym-landau}
\end{equation}

\section{Lattice formalism and computational details}
\label{sec:comput}

In this study, we use the Sheikholeslami--Wohlert fermion action,
\begin{equation}
S_{SW} = S_W - i\frac{a}{4}g_0\csw\sum_x\sum_{\mu\nu} 
 \psibar(x)\smn F_{\mu\nu}(x)\psi(x) \, ,
\label{eq:sw}
\end{equation}
which is on-shell $\order(a)$-improved ($S_W$ is the Wilson action),
along with an off-shell improved\footnote{For full off-shell
improvement, there should also be a gauge dependent improvement term.
The absence of this term will give rise to errors, potentially of
$\order(g^2a)$ \cite{Sharpe:2001cn}.  We assume that this is a small
effect compared with other systematic errors.  Setting this term to
zero is also consistent with mean-field improvement, which is what we
are using in this paper.} quark propagator $S_I$, given by \cquark
\begin{equation}
S_I(x,y) = (1+b_q ma)S_0(x,y) - 2ac'_q\delta(x-y) \, ,
\label{eq:quark-imp}
\end{equation}
where the `unimproved' quark propagator $S_0$ is simply derived from
the inverse of the fermion matrix $M$: $S_0(x,y)=\bra
M^{-1}(x,y;U)\ket$. 

We define the lattice gluon field $A_{\mu}$, which in the continuum
limit becomes $aA_{\mu}^{\text{cont}}$, as
\begin{equation}
\begin{split}
A_\mu(q) & \equiv \sum_x e^{-iq\cdot(x+\muhat/2)}
 A_\mu(x+\muhat/2) \\
 & =  \frac{e^{-iq_{\mu}a/2}}{2ig_0}\left[\left(U_\mu(q)-U^{\dagger}_\mu(-q)\right)
 - \frac{1}{3}\Tr\left(U_\mu(q)-U^{\dagger}_\mu(-q)\right)\right] .
\end{split}
\end{equation}

In order to reduce lattice artefacts, we employ a tree-level
correction scheme, as discussed at length in \cquarks.  The relevant
tree-level expressions, and the definition of the tree-level corrected
vertex form factors, are given in appendix~\ref{sec:tree}.

All the results in this paper have been obtained with the Wilson
gauge action at $\beta=6.0$ on a $16^3\times 48$ lattice.  Using the
hadronic radius $r_0$ \csommer\ to set the scale, this corresponds to
a lattice spacing $a^{-1}=2.12$ GeV.  The gauge
fields were generated with a Hybrid Over-Relaxed algorithm, with
configurations separated by 800 sweeps.
The quark propagators have been generated using a
mean-field improved SW fermion action, with $\csw=1.479$, for one value
of $\kappa=0.1370$, or $ma=0.0579$.  Details of the computation are
given in \cUKQCD.  For the improvement coefficients $b_q$ and
$c'_q$ of (\ref{eq:quark-imp}) we have used the mean-field values
$b_q=1.14, 2c'_q=0.57$.

The gauge fields have been fixed to Landau gauge, using a Fourier
accelerated algorithm \cite{Davies:1988vs} to deal with low-momentum
modes.  The Landau gauge condition has been achieved to an accuracy of
$\frac{1}{VN_C}\sum_{x,\mu}\left|\partial_\mu A_\mu\right|^2 <
10^{-12}$.  Further details of the gauge fixing are given in \cgluon.

For the quark fields, we have used periodic boundary conditions in the
spatial directions and antiperiodic boundary conditions in the time
direction.  Hence, the available momentum values for an $N_i^3\times
N_t$ lattice (with $N_i, N_t$ even numbers and $i=x,y,z$) are
\begin{alignat}{2}
p_i & = \frac{2\pi}{N_i a}\left(n_i - \frac{N_i}{2}\right)
 & ;\qquad
 n_i = & 1,2,\cdots,N_i \; ; \\
p_t & = \frac{2\pi}{N_t a} \left(n_t-\half-\frac{N_t}{2}\right)
& ;\qquad 
 n_t = & 1,2,\cdots,N_t \;.
\label{eq:latt_momenta}
\end{alignat}
For the gluon fields, we have used periodic boundary conditions in all
directions, and thus we have integer momentum values also in the time
direction.  The gluon tensor structure was studied in \cgluon.  Our
data correspond to the `small lattice' in that paper.  There it was
found that 
\begin{equation}
T_{\mu\nu}(0)\sim\operatorname{diag}(1,1,1,1/3) \qquad 
\implies \qquad T(0) \approx \frac{10}{3} \,,
\label{eq:zeromom-tensor}
\end{equation}
where $T_{\mu\nu}$ and $T$ are defined in (\ref{eq:tensor-finvol}).
Significant deviations from the infinite-volume form were also found
for the lowest one or two momentum points used for the symmetric
kinematics.  We have explicitly adjusted these points to account for
this. 
For all other momentum combinations we will be studying here, the
deviation of $T_{\mu\nu}(q)$ from $P^{\text{lat}}_{\mu\nu}(q) =
P_{\mu\nu}(Q(q))$ were found to be negligible.

\section{Determination of $\Zquark$ and $\Zgluon$}
\label{sec:zgluon-zquark}

\FIGURE{
\includegraphics*[width=15cm]{zquark.eps}
\caption{$\Zquark$ as a function of the renormalisation scale $\mu a$,
for $S_0$ (left) and for $S_I$ (right), using the tree-level
correction defined in (\protect\ref{eq:quark-corr}), and without any
momentum cuts . }
\label{fig:zquark-corr}
}
\FIGURE{
\includegraphics*[width=7.0cm]{zgluon.eps}
\caption{$\Zgluon$ as a function of the renormalisation scale $\mu a$,
for 125 configurations.  The line denotes the best fit to the
functional form (\protect\ref{eq:gluon-model}).}
\label{fig:zgluon-all}
}
In order to determine the quark field renormalisation constant
$\Zquark$, we have used the tree-level corrected function $Z(p)$,
defined in (\ref{eq:quark-corr}).
The results, for both $S_0$ (for
which $\zz(p)\equiv1$) and $S_I$ are shown in
figure \ref{fig:zquark-corr}.

In order to determine the gluon field renormalisation constant
$\Zgluon$, we use the simple tree-level correction procedure that was
applied in \cgluon.  The gluon propagator $D(q^2)$ is expressed in terms
of the `lattice momentum' $Q$ [see (\ref{def:lat-Q})], and a `cylinder
cut' is applied to select momenta near the 4-diagonal.  This is shown
as a function of $Qa$ in figure \ref{fig:zgluon-all}.  We have fitted
the gluon propagator to the phenomenological curve (Model A) of
\cgluon,
\begin{gather}
D(Q^2) = Z\left[\frac{AM^{2\alpha}}{(Q^2+M^2)^{1+\alpha}} + 
\frac{1}{Q^2+M^2}
\left[\half\ln\left((Q^2+M^2)(Q^{-2}+M^{-2})\right)\right]^{-d_D}
\right] \, , \label{eq:gluon-model}
\end{gather}
where $d_D=13/22$ is the gluon anomalous dimension.
The best estimates for the parameters are 
\begin{equation}
Z=2.02 ; \qquad A=10.7 ; \qquad M=0.534 ; \qquad \alpha=2.17 \, .
\end{equation}
It should be emphasised that this fit is only performed to facilitate
the computation of the running coupling, and no physical
significance should be attached to the phenomenological parameters
quoted. 

\section{$\lambda_1$ and the running coupling}
\label{sec:vtx-proper}

\subsection{Asymmetric scheme}
\label{sec:asymm}

We have calculated the proper vertex in the asymmetric scheme using
both the un\-im\-prov\-ed quark propagator $S_0$ and the improved propagator
$S_I$. 

We have evaluated $\lambda_1$ by first calculating $H_i(p,q=0)
(i=1,2,3)$ for different values of $p$ (with $p_i=0$), and then used
invariance under the (hy\-per-)cu\-bic group to perform a $Z_3$ average
over $i$ and equivalent values of ${p_\mu}$ (as well as positive and
negative $p_\mu$ values).  But first, we want to verify that the cubic
invariance really holds.  As figure \ref{fig:vtx1234} shows, all the
three spatial components of the (uncorrected) vertex do indeed behave
in the same fashion, within errors.  The discrepancies of the order
$2\sigma$ can be put down to correlations between data at different
momenta, combined with insufficient statistics.

When $p_\mu\ne 0$, $H_\mu(p,q)$ also receives a contribution from
$\lambda_2$. This means that we should not expect $H_4$ to behave
similarly to the other three components, since $p_4=p_t$ is
necessarily non-zero.  The lower panel of figure~\ref{fig:vtx1234}
confirms this --- although part of the difference may also be due to
finite volume effects affecting spatial and time directions
differently.  The form factor $\lambda_2$ will be studied
in a forthcoming paper.
\FIGURE{
\includegraphics*[width=14cm]{vertex_mu.eps}
\caption{Top: $H_1(p,q), H_2(p,q)$ and $H_3(p,q)$ 
for $q=0$, $p=(0,p_t)$, as a
function of $p_t a$; using $S_0$ (left) and $S_I$ (right).
Bottom:$H_4(p,q)$ for $q=0$, $p=(0,p_t)$, as a
function of $p_t a$, for 83 configurations using $S_0$ (left) and for
100 configurations using $S_I$ (right).  Note the different vertical
scales for the upper and lower panels.}
\label{fig:vtx1234}
}

\FIGURE{
\includegraphics*[width=10cm]{lambda1.eps}
\caption{The unrenormalised form factor $\lambda_1(p^2,0,p^2)$ as
a function of $|pa|$, with equivalent momenta averaged.  The form factor
taken from the improved propagator $S_I$ is shown both before and
after tree-level correction.  After tree-level correction, the lattice
data for $S_I$ lie on a single smooth curve.}
\label{fig:lambda1}
}
In figure \ref{fig:lambda1} we show the unrenormalised
form factor $\lambda_1$, obtained by averaging all
$H_i(q=0,p_i=0)$ over equivalent momenta and directions, as a function
of $|pa|$.  As the figure shows, this is a well-defined function of
$p$ (within the statistical errors) for $pa\lesssim1$, both when $S_0$
and $S_I$ is used.  For larger values of $pa$, however, $\lambda_1$
extracted using $S_I$ developes significant ambiguities and a big
`bump' around $pa=1.7$.  This is due to the tree-level
behaviour given in (\ref{eq:vtx-tree-imp-asym}).  Comparing with
figure~1 of \cquark, we see that it is indeed approximately the inverse of
the tree-level quark propagator.  As expected, the irregular behaviour
disappears after tree-level correction, and all the data lie on a
single smooth curve.  For $pa\gtrsim1$ this curve coincides with the
data for $S_0$.

We clearly see a substantial infrared enhancement of $\lambda_1$, in
accordance with the expectations from studies of the gluon and ghost
propagators.  Part of this must be an `abelian' enhancement given by
the QED expression (\ref{eq:bc1}) together with the infrared
suppression of the quark propagator.  We can determine the additional,
`non-abelian' enhancement by plotting the product of $\lambda_1$ and
the quark propagator form factor $Z(p)$, which in QED would be a
momentum-independent constant.  This is shown in
figure~\ref{fig:lambda1byZ}, using the improved propagator $S_I$.  As
we can see, a significant enhancement over and above the `abelian' one
remains, although our lattice is too small to allow us to draw any
further quantitative conclusions.
\FIGURE{
\includegraphics*[width=10cm]{lambda1_by_z.eps}
\caption{The unrenormalised form factor $\lambda_1(p^2,0,p^2)$
multiplied by the quark renormalisation function $Z(p)$, using the
improved propagator $S_I$, as a function of $p$.}
\label{fig:lambda1byZ}
}

\FIGURE{
\vspace{-2mm}
\includegraphics*[width=14cm]{gren.eps}
\caption{$\gmomt(\mu)$ as a function of $\mu$ (GeV), using $S_0$
(left), and $S_I$ (right).}
\label{fig:gren-asym} 
}
Using the values for $\Zquark$ and $\Zgluon$ in section
\ref{sec:zgluon-zquark}, we obtain $\gmomt(\mu)$, which is shown in
figure \ref{fig:gren-asym}, or, equivalently, $\amomt(\mu)$, shown in
figure \ref{fig:alpha}.  The coupling appears to reach a peak at about
1 GeV, below which it drops towards zero.  However, caution is clearly
warranted: the two or three lowest momentum points where this effect
can be observed may well contain substantial finite volume effects
(indeed, this was the case with the gluon propagator on the same
lattice \cgluon), which only a simulation on a larger volume can
resolve.  Thus, at this point, we are not able to tell whether this
may be a finite volume artefact, an artefact of the \MOMT\ scheme, or
a real physical effect.  Similarly, the fact that the peak value
$\amomt\sim0.8$ is very close to typical values for the frozen
coupling extracted from phenomenology \cite{Mattingly:1994ej}, may be
suggestive, but nothing more.
\FIGURE{
\includegraphics*[width=10cm]{alpha.eps}
\caption{$\amomt(\mu)$ as a function of $\mu$ (GeV).  Also shown is the
fit to (\protect\ref{eq:alpha-powcorr}) for $\mu>2.0$ GeV.}
\label{fig:alpha} 
}

Turning now to the ultraviolet behaviour, we attempt to parametrise
the leading nonperturbative and quark mass effects by fitting the
results to the formula \cpow
\begin{equation}
\alpha(\mu) \equiv \frac{g^2(\mu)}{4\pi}
 = \left(1+\frac{c}{\mu^2}\right)\alpha^{\text{2loop}}(\mu) \, ,
\label{eq:alpha-powcorr}
\end{equation}
for $\mu\geq p_{min}$, where $\alpha^{\text{2loop}}$ is the two-loop
running coupling,
\begin{equation}
4\pi\alpha^{\text{2loop}}(\mu) = \frac{1}{b_0\ln(\mu^2/\Lambda^2)
+\frac{b_1}{b_0}\ln\ln(\mu^2/\Lambda^2)} \, ,
\label{eq:gsqr-2loop}
\end{equation}
with $b_0=11/16\pi^2$ and $b_1=102/(16\pi^2)^2$ the leading
coefficients of the $\beta$-function, for varying values of $p_{min}$.
The results are shown in table~\ref{tab:fitparams}.  In all cases, the
numbers obtained using $S_0$ are almost identical to those obtained
using $S_I$, so we only report the latter.
\TABLE{
\begin{tabular}{llrrrr}
$p_{min}$ (GeV) & $n$ & $\Lambda^0$ (MeV) & $c (\text{GeV}^2)$
 & $\Lambda$ (MeV) & $\Lambda^r$ (MeV) \\ \hline
2.00 & 155 & 582\err{63}{128}  &  1.4\err{2.4}{1.5} & 382\err{192}{232} & 425\err{216}{259} \\
2.25 & 149 & 567\err{69}{135}  &  1.1\err{2.6}{1.6} & 407\err{208}{250} & 454\err{236}{276} \\
2.50 & 139 & 555\err{79}{141}  &  0.6\err{2.9}{1.8} & 445\err{223}{272} & 497\err{253}{305} \\
2.75 & 125 & 544\err{97}{157}  &  0.3\err{3.3}{2.3} & 461\err{275}{304} & 515\err{312}{341} \\
3.00 & 112 & 534\err{103}{169} & -0.5\err{3.6}{2.4} & 523\err{331}{372} & 586\err{376}{415} \\
3.25 & 101 & 528\err{114}{175} & -1.7\err{4.2}{2.4} & 613\err{366}{410} & 686\err{411}{456} \\
3.50 &  85 & 536\err{124}{182} & -2.2\err{4.2}{2.4} & 662\err{386}{428} & 740\err{434}{480} 
\\ \hline
\end{tabular}
\caption{Fit parameters, using different momentum ranges.  $p_{min}$
denotes the lower end of the fit window; the maximum in all cases
being the maximum total momentum 5.75 GeV. $n$ is the number of
momentum points used in the fit.  $\Lambda^0$ is the value
obtained for $\Lmomt$ without power correction, using
(\protect\ref{eq:Lambda_QCD}), while $\Lambda^r$ is the value obtained
by using the fitted value for $c$ to extract $\alpha^{\text{2loop}}$
and feeding this into (\protect\ref{eq:Lambda_QCD}).}
\label{tab:fitparams}
}
We may also, if we ignore the power corrections (i.e., set $c=0$),
compute $\Lmomt$ directly according to the inverse of
(\ref{eq:gsqr-2loop}), 
\begin{equation}
\Lambda = \mu e^{-\frac{1}{2b_0g^2(\mu)}}
\left(b_0g^2(\mu)\right)^{-\frac{b1}{2b_0^2}} \, .
\label{eq:Lambda_QCD}
\end{equation}
The results of this are shown in figure
\ref{fig:Lambda-asym}.  The numbers obtained by fitting this to a
constant above $p_{min}$ are also reported in
table~\ref{tab:fitparams}.  These numbers are consistent with the result of
fitting $\alpha(\mu)$ to (\ref{eq:gsqr-2loop}).  We may also repeat
this procedure after absorbing the power correction
(\ref{eq:alpha-powcorr}) in our definition of $g_R$, using the value
for $c$ from our fit.  The result of this is also shown
in figure \ref{fig:Lambda-asym} and reported in table
\ref{tab:fitparams}.
\FIGURE{
\vspace{-3mm}
\includegraphics*[width=14cm]{Lambda_asym.eps}
\caption{$\Lmomt(\mu)$ (GeV) as a function of $\mu$ (GeV).  Left:
without power correction.  Right: including the power correction of
(\protect\ref{eq:alpha-powcorr}) from a fit to $\mu>2.0$ GeV.  The
lines indicate the preferred value for $\Lmomt$, with a 67\%
confidence interval.}
\label{fig:Lambda-asym} 
}

In lattice studies of momentum-space quantities, the momentum variable
is to some extent arbitrary.  We may choose any of the variables $p$,
$K(p)$, $Q(p)$, $\Kt(p)$ or any other variable as long as it
approaches $p$ in the infrared and in the continuum limit, i.e.\ for
$pa\ll1$.  If the continuum tree-level form of the quantity is
momentum-dependent, we may use this to guide our choice of variable
\cite{Bonnet:2000kw}; however, when it is not, the choice remains
largely arbitrary \cite{Bonnet:2002ih}.  Since the tree-level
continuum vertex is momentum-indepentent, this is the situation we
find ourselves in here.  In order to quantify the resulting ambiguity,
we have, in addition to the `na\"{\i}ve' momentum $p$, performed fits
using $K(p)\equiv\sqrt{\sum_\mu\sin^2(ap_\mu)}/a$, which appears in the tree-level lattice vertex
(\ref{eq:vtx-tree}), as well as $K_z(p)\equiv K(p)/\zz(p)$, which is
the momentum variable that makes the tree-level quark propagator take
its continuum form.  The use of this variable may be justified because
the correction factor $\zz(p)$ appears also in the tree-level vertex,
and also from the Ball--Chiu relation (\ref{eq:bc1}).  The results of
the fits are given in tables \ref{tab:fits-k} and \ref{tab:fits-kz}.
\TABLE{
\begin{tabular}{llrrrr}
$K_{min}$ (GeV) & $n$ & $\Lambda^0$ (MeV) & $c (\text{GeV}^2)$
 & $\Lambda$ (MeV) & $\Lambda^r$ (MeV) \\ \hline
2.00 & 146 & 436\err{59}{109}  &  0.4\err{2.2}{1.3} & 343\err{209}{221} & 374\err{236}{242} \\
2.25 & 123 & 423\err{76}{121}  & -0.2\err{2.2}{1.3} & 404\err{246}{268} & 443\err{275}{296} \\
2.50 & 101 & 423\err{90}{130}  & -0.2\err{3.3}{1.7} & 400\err{303}{309} & 437\err{342}{340} \\
2.75 &  78 & 421\err{98}{141}  & -0.8\err{3.9}{1.8} & 464\err{336}{353} & 511\err{376}{390} \\
3.00 &  46 & 425\err{130}{168} & -3.0\err{3.0}{1.6} & 725\err{391}{463} & 805\err{436}{518} \\
3.25 &  24 & 455\err{161}{190} & -2.8\err{4.0}{2.4} & 699\err{496}{544} & 780\err{553}{609} \\
3.50 &   9 & 470\err{189}{233} &  0.0\err{21.5}{5.5} & 421\err{813}{419} & 468\err{910}{465} 
\\ \hline
\end{tabular}
\caption{As table \protect\ref{tab:fitparams}, using $K(p)$ as our
momentum variable.  The maximum available momentum here is 3.70 GeV.}
\label{tab:fits-k}
}
\TABLE{
\begin{tabular}{llrrrr}
$K_{z,min}$ (GeV) & $n$ & $\Lambda^0$ (MeV) & $c (\text{GeV}^2)$
 & $\Lambda$ (MeV) & $\Lambda^r$ (MeV) \\ \hline
2.00 & 155 & 615\err{65}{134}  &  1.1\err{2.0}{1.3} & 442\err{192}{233} & 483\err{218}{259} \\
2.25 & 150 & 605\err{76}{145}  &  1.1\err{2.1}{1.5} & 443\err{202}{241} & 484\err{225}{266} \\
2.50 & 145 & 596\err{83}{149}  &  0.8\err{2.1}{1.5} & 461\err{194}{258} & 503\err{223}{283} \\
2.75 & 133 & 591\err{94}{164}  &  1.3\err{2.6}{1.7} & 429\err{217}{257} & 466\err{254}{283} \\
3.00 & 121 & 583\err{106}{172} &  1.9\err{3.2}{2.5} & 395\err{227}{257} & 430\err{257}{281} \\
3.25 & 112 & 578\err{110}{177} &  2.2\err{4.3}{2.8} & 377\err{237}{237} & 410\err{268}{260} \\
3.50 & 102 & 571\err{116}{184} &  2.4\err{5.0}{3.3} & 369\err{317}{257} & 402\err{356}{271} 
\\ \hline
\end{tabular}
\caption{As table \protect\ref{tab:fitparams}, using $K_z(p)$ as our
momentum variable.  The maximum available momentum here is 5.05 GeV.}
\label{tab:fits-kz}
}

From figure \ref{fig:alpha} and the right-hand panel of
figure~\ref{fig:Lambda-asym} it would appear that the data are very well
represented by a power-corrected two-loop running coupling as in
(\ref{eq:alpha-powcorr}), all the way down to 1.5 GeV if not lower.
However, a glance at tables \ref{tab:fitparams}--\ref{tab:fits-kz}
reveals several problems with this.

Firstly, the fits are nowhere near stable.  As the starting point for
the fits goes from 2.5 to 3.5 GeV, the best value for $\Lambda$
increases by 50\% when using $p$ as our momentum variable,
and the power correction goes from positive to negative.  Secondly,
the `refitted' value for $\Lambda$, although always perfectly
consistent with that obtained from (\ref{eq:alpha-powcorr}), is
consistently about 10\% higher.

Thirdly, the fit values depend critically on which momentum value is
used.  To some extent this is simply because the values of $p$,
$K(p)$, and $K_z(p)$ may be very different when $pa\gg1$, so different
data are included in the fits.  This is reflected in the different
number of points for the same numerical value of the starting
momentum.  However, it also reflects a deeper ambiguity due to the
finite lattice spacing.  I.e., although there is no violation of O(4)
symmetry or other obvious signs of lattice artefacts in our data, and
the near-perfect agreement between the $S_0$ and $S_I$ results may be
taken as an indication that lattice spacing errors are very small,
those errors that do persist make a determination of a sensitive
quantity such as $\Lmomt$ prone to large uncertainties.  It should be
noted that we observe considerable anisotropy in the high-momentum
region when $g_R$ is plotted as a function of $K(p)$ or $K_z(p)$ as
opposed to $p$, indicating that these are not the appropriate momentum
variables in this case.  Only by repeating the simulation at a smaller
lattice spacing can this issue be properly resolved, however.

Taking all this into account, we take as our best estimate for
$\Lambda$ the average of all the fits starting from 3.0 GeV (both with
and without the power correction).  This gives
$\Lmomt=530\err{260}{320}\pm100\pm50$ MeV, where the first set of errors are
statistical, the second are due to the ambiguities in the choice of
momentum variable, and the third is the intrinsic 10\% systematic
uncertainty in the lattice spacing in the quenched approximation.

\subsection{Symmetric scheme}
\label{sec:symm}

At the symmetric point $2p+q=0$ we use only the improved propagator
$S_I$, and thus the form factor $\lambda'_1$ receives substantial
tree-level correction according to
(\ref{eq:tree-imp-sym})--(\ref{eq:lat-sym-trans}).  The tree-level
corrected result is shown in figure~\ref{fig:lambda1-symm}.
In order to reduce the statistical noise, we have averaged data for
nearby momenta, within $\Delta pa<0.05$. 
We see that the data ara still considerably more noisy than for the asymmetric
$\lambda_1$ of figure~\ref{fig:lambda1}, but exhibit qualitatively
the same behaviour.  It appears that $\lambda'_1(p^2,4p^2,p^2)$ is more
strongly infrared enhanced than $\lambda_1(p^2,0,p^2)$, but the noise
makes it difficult to draw any definite conclusion.
\FIGURE[t]{
\centering
\includegraphics*[width=9cm]{lambda1_sym_smear.eps}
\caption{The unrenormalised, tree-level corrected form factor
$\lambda'_1(p^2,4p^2,p^2)$ as a function of the quark momentum $|pa|$,
from 498 configurations.}
\label{fig:lambda1-symm}
}

\FIGURE{
\includegraphics*[width=9cm]{gren_sym_smear.eps}
\caption{$\gmomb(\mu)$ as a function of $\mu$ (GeV).}
\label{fig:gren-symm} 
}
The \MOMB\ running coupling $\gmomb(\mu)$ is shown as a function of
the renormalisation scale $\mu$ in figure \ref{fig:gren-symm}.
The most obvious difference from the \MOMT\ coupling of figure
\ref{fig:gren-asym} is that the noise is far worse, and we are not
able to get any signal for $\Lambda$ from these data.

\section{Matching to \MSB}
\label{sec:match}

The relation between the scale parameters in two renormalisation
schemes A and B is given by \cite{Celmaster:1979km}
\begin{equation}
\frac{\Lambda_A}{\Lambda_B} = \exp\biggl[-\frac{C_{AB}}{2b_0}\biggr]
\, ,
\label{eq:match-Lambda}
\end{equation}
where ${C_{AB}}$ is the one-loop coefficient in the expansion of the
coupling $g_B^2$ in terms of $g_A^2$.

The complete one-loop expressions for the relevant form factors in the
two kinematics we are studying, are given in
appendix~\ref{sec:one-loop}.  Here we only reproduce the results for
the running coupling in Landau gauge, for $N_f=0$.  The \MOMT\
coupling is given by
\begin{multline}
\gmomt(\mu) = 
\gms(\mu)\Biggl[1+\biggl(\frac{151}{24}-\frac{3}{4}\frac{m^2}{\mu^2}
 - \frac{9}{4}\ln\Bigl(1+\frac{m^2}{\mu^2}\Bigr) \\
 + \frac{m^2}{\mu^2}\ln\Bigl(1+\frac{\mu^2}{m^2}\Bigr)
    \Bigl[\frac{4}{3} + \frac{25}{12}\frac{m^2}{\mu^2}\Bigr]\biggr)
 \frac{\gms^2(\mu)}{16\pi^2} + \order(g^4)\Biggr] \, .
\label{eq:gmomt-ms}
\end{multline}
However, at asymptotically large momenta, where one-loop perturbation
theory becomes valid, the corrections due to the mass term can be
ignored.  From (\ref{eq:match-Lambda}) we thus find that
\begin{equation}
\frac{\Lmomt}{\Lms} = \exp\frac{151}{264} = 1.77 \, .
\label{eq:Lambdamom-ms}
\end{equation}
Using our `best value' for $\Lmomt$, we obtain
\begin{equation}
\Lms^{N_f=0} = 300\err{150}{180}\pm55\pm30 \text{MeV}.
\end{equation}
These numbers are above those obtained by other methods \cLambdas,
which yield a
`world average' of $\Lms^{N_f=0}=240(10)$ MeV.  With our large
statistical and systematic uncertainties, our value is however fully
consistent with the `world average'.

The \MOMB\ coupling in the massless limit is
\begin{equation}
\gmomb(\mu) = \gms(\mu)\Biggl\{
1+\bigg(\frac{4}{9}\ln2+\frac{793}{72}\bigg)\,
                 \frac{\gms(\mu)}{16\pi^2}+\order(g^4)
\Bigg\} \, ,
\end{equation}
which gives for $\Lmomb$
\begin{equation}
\frac{\Lmomb}{\Lms} = \exp\left(\frac{8\ln2/9 + 793/36}{22}\right)
 = 2.80 \, .
\end{equation}
If, instead, we renormalise the vertex at the gluon momentum, we find
\begin{equation}
\frac{\Lmomb^g}{\Lms} = \exp\left(\frac{89\ln2/9 + 793/36}{22}\right)
 = 3.72 \, .
\end{equation}

\section{Discussion and outlook}
\label{sec:outlook}

We have studied the quark-gluon vertex in the Landau gauge, in
quenched QCD with $\order(a)$-improved Wilson fermions, at two
different kinematical points: an `asymmetric' point, where the gluon
momentum $q$ is zero, and a `symmetric' point, where $q=-2p$, in other
words
the incoming quark has equal and opposite momentum to the outgoing
quark.

We have focused on the form factor $\lambda_1$, which is proportional
to the running coupling and, in the decomposition given by
(\ref{eq:long-components}), (\ref{eq:trans-components}), is the only
form factor that is expected to be ultraviolet divergent.  At the symmetric point, we
are unable to study this form factor directly, and examine instead the
linear combination $\lambda'_1\equiv\lambda_1-q^2\tau_3$.  We observe
that in both kinematics, $\lambda_1 (\lambda'_1)$ is substantially
enhanced in the infrared.  At the asymmetric point, this enhancement
is significantly stronger than that expected in QED due to the
well-established enhancement of the quark propagator form factor
$A(p)$.  At the asymmetric point, no such direct comparison with QED
is possible due to the admixture of $\tau_3$, which is left
unconstrained by the Ward--Takahashi (or Slavnov--Taylor) identity.
However, the qualitative picture is the same.

The lattice volume in this study is relatively small (a spatial length
of $\sim$1.5 fm and a total volume of 15--16 fm${}^4$), so the
infrared behaviour may well be contaminated by substantial finite
volume effects.  Although we have explicitly accounted for the large
finite-volume effects appearing in the tensor structure of the gluon
propagator, we have no guarantee that there are not substantial
residual effects that, at the asymmetric point, could play an important role
for all momenta.
However, excellent agreement with results for $\Lms$ obtained by other
methods have been obtained from the three-gluon vertex in a \MOMT\
scheme on symmetric lattices \cggg, and there appears to be no reason
why the situation should be much worse in our case.  The qualitative
similarity between the symmetric and asymmetric point might also be
taken as an indication that finite volume effects, although possibly
sizeable, do not dominate.  In any case, it would be desirable to
perform the simulation on a larger lattice in order to have a better
resolution of the momentum in directions other than the time
direction.  This is also the only way we would be able to settle the
issue of whether the running coupling is frozen, or possibly goes to
zero.

We have used the results for $\lambda_1$ at the asymmetric point to
determine the running coupling $\alpha_s$ in a zero-momentum (\MOMT)
renormalisation scheme, and obtained from this a nonperturbative
estimate of $\Lms^{N_f=0}$.  Our
main results are for the strong coupling 
$\amomt^{N_f=0}(2\text{GeV})=0.36(4);\,
\ams^{N_f=0}(2\text{GeV})=0.28(3)$, and for the QCD scale
$\Lms^{N_f=0}=300\err{150}{180}\pm55\pm30$ MeV, where the first set of errors
are statistical, the second due to ambiguities in defining the
momentum variable, and the third due to uncertainty in the lattice
spacing.  This is consistent with, although slightly higher than other
estimates for $\Lms$.

Although the excellent agreement between the results using the `unimproved'
propagator $S_0$ and the `improved' propagator $S_I$ indicate that our
tree-level correction scheme has successfully accounted for the large
high-momentum lattice artefacts, and that
residual  $\order(a)$ errors are not a significant factor, the need
for large tree-level corrections still implies some uncertainty about
the results, at least in the intermediate momentum regime.  A fermion
action which is more well-behaved at high momenta, such as overlap
fermions, would be a great improvement.

The main source of systematic uncertainty, and of possible
discrepancies between our result for $\Lms^{N_f=0}$ and those of other
determinations, is that we have not been able to access sufficiently
high momenta, where two-loop scaling should be valid, nor have we
taken into account higher-order perturbative effects, which should
extend the range of validity for the perturbative matching.
Experience from the 3-gluon vertex \cggg\ suggests that both a large
momentum window and 3-, perhaps 4-loop running of the $\beta$-function
are needed to obtain reliable results.  This requires simulations at
smaller lattice spacings, as well as a two-loop calculation of the
$\lambda_1$ form factor in the relevant kinematical limit.  Both are
computationally very expensive.

As we mentioned in the introduction, two-loop calculations have
already been performed in both an asymmetric \cite{Chetyrkin:2000dq}
and a symmetric \cite{Chetyrkin:2000fd} kinematics.  Neither is,
however, the kinematics we are employing here. 

In the \MOMB\ (symmetric) kinematics, we have been unable to get a
reasonable signal for the running coupling.  The main reason for this
is statistical noise, but the need for large tree-level corrections
is clearly also a significant factor.  For this reason it would be
essential, if we were to attempt a more accurate determination of the
vertex in this kinematics, to choose a fermion discretisation which is
not afflicted by such problems.

In a precision study, the quark mass must also be handled carefully.
Here, we have merely included the quark mass in the overall power
correction, which has been determined numerically.  An obvious next
step would be to study the vertex at a second quark mass.  It would be
an advantage, also for this purpose, to use a fermion action which
respects chiral symmetry, such as overlap fermions, or a remnant
thereof, such as staggered fermions.

The large numerical uncertainties have prevented us from obtaining any
reliable estimate of the power correction.  An alternative approach
would be to calculate analytically the size of the power corrections
from the condensates involved, of which the dominant is expected to be
the chiral condensate, using the available estimates for the values of
the condensates.

The running coupling extracted in the \MOMT\ scheme reaches a maximum
at 0.8 GeV.  A similar result was found for the three-gluon vertex in
an analogous \MOMT\ scheme \cgggs.  This would correspond to a zero
in the $\beta$-function at the maximum coupling, with double values
below that.  It has been suggested \cite{Alkofer:1999gd} that this can
be related to infrared singularities in the ghost self-energy.  Such
singularities should not affect symmetric momentum subtraction
schemes.  Our results in the \MOMB\ scheme can neither confirm nor
refute this conjecture.

In order to resolve this issue, and to pin down the low- and
intermediate-momentum behaviour of the quark--gluon vertex,
simulations on larger, and possibly coarser lattices are necessary.
This is an orthogonal line of inquiry to that needed to determine the
running coupling along with the power corrections, which requires much
finer, but not larger lattices.

Work is currently in progress to determine the form factors
$\lambda_2$ and $\lambda_3$ at the asymmetric ($q=0$) point.  These
form factors both vanish at tree level in the continuum, but must be
non-zero nonperturbatively in order to fulfil the Slavnov--Taylor
identity.  A complete determination of all form factors would be a
natural next step.  This is, however, not possible in Landau gauge
because of the transversality condition.  For this reason, and also
because the gauge dependence of the vertex is in itself of theoretical
importance, it would be of great interest to study the vertex in a
generic covariant gauge \cite{Giusti:1997kf,Giusti:1999im}.  This
would also allow calculations in the unmodified symmetric \MOMB\
scheme, which might have some advantages over the modified scheme we
have used here.

At present, there is no known method to reliably assess the effect of
Gribov copies.  All numerical methods founder on the fact that as the
physical volume increases, the number of Gribov copies also increases,
and it becomes impossible to ascertain that one has found either the
absolute maximum of the gauge fixing functional, or any other unique
representative.  Choosing a gauge without Gribov copies, such as the
Laplacian gauge \cite{Vink:1992ys} or axial gauges, does not solve the
problem, since results in one gauge tell us nothing about the effect
of Gribov copies in a different gauge.

On a practical level, attempting to select, however imperfectly, the
absolute maximum, using e.g.\ `brute force' \cite{Cucchieri:1997dx},
simulated annealing \cite{Langfeld:2001cz}, or smeared gauge fixing
\cite{Hetrick:1997yy} is in principle 
worthwhile.  At present, however, we would expect any signal showing a
difference between the na\"{\i}ve (minimal) Landau gauge and the
fundamental modular domain to be swamped by statistical noise for
three-point functions such as the quark--gluon vertex.

\section*{Acknowledgments} 

We wish to give special thanks to Tony Williams for numerous long and
productive discussions 
and a careful reading of the manuscript.  We also thank Claudio
Parrinello, Derek Leinweber, Reinhard Alkofer, 
Peter Watson and Olivier P\`ene for stimulating discussions and
advice.  We acknowledge
support from the Norwegian Research
Council, the Australian Research Council, the European Union TMR
network ``Finite temperature phase transitions in particle physics'',
and FOM (The Netherlands).
This study was performed using UKQCD configurations, 
produced on a Cray T3D based at EPCC, University of Edinburgh, using
UKQCD Collaboration CPU time under PPARC Grant GR/K41663.

\appendix
\section{Tensor decomposition of the vertex}
\label{sec:decompose}

The Lorentz structure of the vertex in the continuum consists of 12
independent vectors and can be written as
\begin{equation}
\Lambda_\mu \equiv -ig_0\Gamma_\mu = -ig_0\sum_{i=1}^{12}f_i F_\mu^i
\,, 
\end{equation}
where
\begin{alignat}{4}
F_\mu^1 &= p_\mu & ;\qquad F_\mu^2 &= q_\mu &;\qquad F_\mu^3 &=
\gamma_\mu &;\qquad F_\mu^4 &= \pslash p_\mu \, ; \notag \\
F_\mu^5 &= \pslash q_\mu &;\qquad F_\mu^6 &= \qslash p_\mu &;\qquad
F_\mu^7 &= \qslash q_\mu &;\qquad F_\mu^8 &= \pslash\gamma_\mu \, ; \\
F_\mu^9 &= \qslash\gamma_\mu\, &;\qquad F_\mu^{10} &= \pslash\qslash
p_\mu \,&;\qquad F_\mu^{11} &= \pslash\qslash q_\mu \,&;\qquad F_\mu^{12} &=
\pslash\qslash\gamma_\mu \, . \notag
\end{alignat}
It is useful to divide the vertex into a `Slavnov--Taylor'
(non-transverse) part and a transverse part, as is commonly done in QED:
\begin{equation} 
\Lambda_\mu(p,q) = 
 \Lambda_\mu^{(ST)}(p,q) + \Lambda_\mu^{(T)}(p,q) \, .
\end{equation}
The ST part is that part of the vertex that saturates the
Slavnov--Taylor identity (\ref{eq:sti}) and contains no kinematical
singularities.  It is often, misleadingly, referred to as the
`longitudinal' part, although it also contains a transverse component.

We will make use of the QED decomposition
\cite{Davydychev:2000rt,Ball:1980ay,Kizilersu:1995iz} of the
fermion--gauge-boson vertex function, which is usually given in
Minkowski space.  We wish to write the euclidean-space equivalent, in
such a way that all the scalar form factors $\lambda_i$ and $\tau_i$
are the same as in Minkowski space.  The usual procedure is to apply
the Wick rotation ($p_0\to ip_4, p_i\to-p_i, \gamma_0\to\gamma_4,
\gamma_i\to i\gamma_i$), but since the vertex is a four-vector, this
is not a linear transformation in our case.  Our prescription is to
create a Lorentz scalar by contracting the vertex with $\gamma_\mu$
and require that the Wick-rotated Minkowski result is identical to
what we obtain by performing this operation in euclidean space.  In
particular, any euclidean scalar function should be equal to the
Minkowskian scalar function sampled at spacelike (i.e.\ negative)
momenta:
\begin{equation}
f_E(p^2,q^2,k^2) \equiv f_M(-p^2,-q^2,-k^2) \, ,
\end{equation}
where $f$ can be $\tau_i$ or $\lambda_i$.

Following this prescription, we can write the ST part as [with $k=p+q$]
\begin{equation}
\Lambda_\mu^{(ST)}(p,q)=
 -ig\sum_{i=1}^{4}\lambda_i(p^2,q^2,k^2)L_{i,\mu}(p,q) \, ,
\label{eq:longitudinal}
\end{equation}
where the euclidean-space functions $L_{i,\mu}$ are given by
\begin{alignat}{2}
L_{1,\mu} & = \gamma_\mu & ;\qquad
L_{2,\mu} & = -(2\pslash+\qslash) (2p+q)_\mu  \,; \notag \\
L_{3,\mu} & = -i(2p+q)_\mu & ;\qquad
L_{4,\mu} & = -i\sigma_{\mu\nu}(2p+q)_\nu \, .
\label{eq:long-components}
\end{alignat}

For the purely transverse part $\Lambda^{(T)}$, we will use the decomposition of
\cite{Kizilersu:1995iz}, which differs slightly from the one of
\cite{Davydychev:2000rt,Ball:1980ay}.  This decomposition is
preferable because it is free of kinematical singularities in all
covariant gauges.  Moreover, as we shall see, the relation between the
ST and purely transverse parts of the vertex becomes more transparent
in this basis.  The purely transverse part of the vertex  is specified by
$q_\mu\Lambda^{(T)}_\mu(p,q)=0$ and satisfies $\Lambda^{(T)}_\mu(p,0)=0$,
and we write
\begin{equation}
\Lambda_\mu^{(T)}(p,q) =
 -ig\sum_{i=1}^{8}\tau_i(p^2,q^2,k^2)T_{i,\mu}(p,q) \, ,
\label{eq:transverse}
\end{equation}
where the euclidean-space functions $T_i$ are given by
\begin{align}
\begin{split}
T_{1,\mu} & =  i\left[p_\mu q^2-q_\mu(p\cdot q)\right] \, ; \\
T_{2,\mu} & =  \left[p_\mu q^2-q_\mu(p\cdot q)\right](2\pslash+\qslash) \, ; \\
T_{3,\mu} & = \qslash q_\mu - q^2\gamma_\mu \, ; \\
T_{4,\mu} & = -i\left[q^2\smn(2p+q)_\nu + 2q_\mu\snl p_\nu q_\lambda\right] \, ; \\
T_{5,\mu} & = -i\sigma_{\mu\nu}q_\nu \, ; \\
T_{6,\mu} & =  q\cdot(2p+q)\gamma_\mu - \qslash(2p+q)_\mu \, ; \\
T_{7,\mu} & =  \frac{i}{2} q\cdot(2p+q)\left[(2\pslash+\qslash)\gamma_\mu -
  (2p+q)_\mu\right] - i(2p+q)_\mu\sigma_{\nu\lambda}p_\nu q_\lambda \, ; \\
T_{8,\mu} & = - \gamma_\mu\sigma_{\nu\lambda}p_\nu q_\lambda
 - \pslash q_\mu + \qslash p_\mu \,.
\label{eq:trans-components}
\end{split}
\end{align}
Charge conjugation symmetry dictates that all the $\lambda_i$'s and
$\tau_i$'s are even with respect to interchanges of $p^2$ and $k^2$
(or $(p+q)^2$), except for $\lambda_4, \tau_4$ and $\tau_6$, which are odd.

In this decomposition, the transverse projection of the ST part of the
vertex is given by
\begin{align}
P_{\mu\nu}(q)L_{1,\nu}(p,q) &= -\frac{1}{q^2}T_{3,\mu}(p,q) \, ; \\
P_{\mu\nu}(q)L_{2,\nu}(p,q) &= -\frac{2}{q^2}T_{2,\mu}(p,q) \, ; \\
P_{\mu\nu}(q)L_{3,\nu}(p,q) &= -\frac{2}{q^2}T_{1,\mu}(p,q) \, ; \\
P_{\mu\nu}(q)L_{4,\nu}(p,q) &= \phantom{-}\frac{1}{q^2}T_{4,\mu}(p,q) \,.
\end{align}
Thus, we will define the following modified form factors, which
appear in the transverse-projected vertex:
\begin{alignat}{2}
\lambda'_1 &= \lambda_1 - q^2\tau_3 &\, ; \qquad 
\lambda'_2 &= \lambda_2 - \frac{q^2}{2}\tau_2 \, ;\\
\lambda'_3 &= \lambda_3 - \frac{q^2}{2}\tau_1 & \, ;\qquad
\lambda'_4 &= \lambda_4 + q^2\tau_4 \, .\notag
\end{alignat}

\section{Tree-level lattice expressions}
\label{sec:tree}

We define and use the following momentum variables, which
may be used to bring the lattice tree-level expressions into a more
continuum-like form,
\begin{align}
K_\mu(p) & \equiv  \frac{1}{a}\sin(p_\mu a) \label{def:lat-K} \,, \\
Q_\mu(p) &
 \equiv  \frac{2}{a}\sin(p_{\mu}a/2) = \frac{\sqrt{2}}{a}\sqrt{1-\cos(p_\mu a)} 
\label{def:lat-Q} \,, \\
\Kt_\mu(p) & \equiv \half K_\mu(2p) = \frac{1}{2a}\sin(2p_\mu a)
\label{def:lat-Kt} \,, \\
C_\mu(p) & \equiv \cos(p_\mu a) \,.\label{def:lat-C}
\end{align}

At tree level, the dimensionless momentum-space propagator $S_0(p)$ is
identical to the free Wilson propagator,
\begin{equation}
S_0^{(0)}(p) = \frac{-ia\kslash(p) + ma + \half a^2 Q^2(p)}
{a^2 K^2(p) + \left(ma+\half a^2 Q^2(p)\right)^2} \, .
\label{eq:free-wilson}
\end{equation}
The tree-level form of the $\order(a)$-improved propagator $S_I$ is
given by
\begin{equation}
S_I^{(0)}(p) = (1+b_q am)S_0^{(0)}(p) - 2ac'_q  
 = \frac{\zz(p)}{ia\kslash(p) + am \zmz(p)} \, ,
\label{eq:qprop-tree-imp}
\end{equation}
where
\begin{align}
\zz(p) &= \frac{1+\bma}{D_I(p)}
 \Bigl[ K^2(p) + \bigl(m+\half Q^2(p)\bigr)^2 \Bigr] \, , \\
\intertext{with}
D_I(p) &= (1+b_q am)^2K^2(p) + B_I^2(p) \, , \\
B_I(p) &= (1+b_q am)\bigl(m+\half Q^2(p)\bigr)
 - 2ac'_q\Bigl[ K^2(p) + \bigl(m+\half Q^2(p)\bigr)^2 \Bigr] \, .
\end{align}
The tree-level functions $\zz$ and $\zmz$ (which is given in
\cquarkII\ and will not be reproduced here) give
rise to very large finite-$a$ effects at intermediate and large
momenta.  To reduce these lattice artefacts, and bring the
high-momentum behaviour of the quark propagator into contact with the
continuum perturbative behaviour, we employ the tree-level correction
scheme defined in \cquarks, where the quark propagator is written as
\begin{equation}
  S^{-1}(p) = \frac{1}{Z(p)\zz(p)} 
  \left[ia\kslash(p) + aM^h(p)\zmp(p) + a\dmm(p) \right] ,
\label{eq:quark-corr}
\end{equation}
with the functions $\zz,\zmp$ and $\dmm$ denoting the tree-level
behaviour.  We call the functions $Z(p)$ and $M^h(p)$ the tree-level
corrected quark form factors.  Here, we are only interested in $Z(p)$,
which can be related to the quark field renormalisation constant
$\Zquark$, so we can ignore the mass correction functions $\zmp,\dmm$
which are defined in \cquarkII.

At tree level, the Landau gauge gluon propagator becomes
\begin{equation}
D_{\mu\nu}(q) = P^{\text{lat}}_{\mu\nu}(q)D(Q^2)
 = \left(\delta_{\mu\nu}-\frac{Q_\mu(q)Q_\nu(q)}{Q^2(q)}\right)
    \frac{1}{Q^2(q)} \, .
\label{eq:gluon-tree}
\end{equation}
In this notation, the gluon propagator requires no further tree-level
correction. 

The tree-level lattice vertex using the `unimproved' propagator $S_0$
is \cite{Heatlie:1991kg,Capitani:1995qn}
\begin{multline}
\Lambda^{a(0)}_{0,\mu}(p,q) = -ig_0 t^a\Biggl( \gamma_\mu
\cos\frac{a(2p+q)_\mu}{2} - i\sin\frac{a(2p+q)_\mu}{2} \\
 - i\frac{\csw}{2}\sum_\nu\smn\cos\frac{aq_\mu}{2}\sin aq_\nu\Biggr) \, .
\label{eq:vtx-tree}
\end{multline}
The constant term $c_q'$ in $S_I$ does not contribute to the
unamputated vertex $V_\mu$ in (\ref{def:vtx}), since $\bra A_\mu\ket=0$.
Thus, the improved vertex at tree level is given by
\begin{equation}
\Lambda^{a(0)}_{I,\mu}(p,q) =
\bmap S_I^\z(p)^{-1}S_0^\z(p)\Lambda^{a(0)}_{0,\mu}(p,q) 
S_0^\z(p+q)S_I^\z(p+q)^{-1} \, .
\label{eq:tree-imp}
\end{equation}
The full expression is very complicated, but it simplifies greatly for
the two cases (symmetric and asymmetric) in which we are interested.

\subsection{Asymmetric kinematics}
\label{sec:tree-asym}

In this case the gluon momentum $q=0$, while the quark momentum is
`orthogonal' to the vertex, i.e.\ the $\mu$-component $p_\mu$ of the
quark momentum is zero.  Then the tree-level lattice vertex
(\ref{eq:vtx-tree}) reduces to
\begin{equation}
\Lambda_{0,\mu}^{a(0)}(p,0)=-ig_0t^a\gamma_{\mu} \, .
\end{equation}
Making use of this, along with the unimproved (\ref{eq:free-wilson})
and improved (\ref{eq:qprop-tree-imp}) propagators, the improved
vertex (\ref{eq:tree-imp}) becomes
\begin{equation}
\begin{split}
\Lambda^{a(0)}_{I,\mu}(p,0)|_{p_\mu=0} & =
(1+b_qam)S_I^{(0)}(p)^{-1}S_0^{(0)}(p)
\Lambda_{0,\mu}^{a(0)}(p,0)S_0^{(0)}(p)S_I^{(0)}(p)^{-1}\\
 & = -ig_0 t^a\gamma_\mu/Z^{(0)}(p) \, .
\label{eq:vtx-tree-imp-asym}
\end{split}
\end{equation}
Thus, at this point, the tree-level corrected vertex may be defined
according to
\begin{equation}
\Lambda^a_{I,\mu}(p,0)|_{p_\mu=0} =
-it^a\frac{1}{Z^{(0)}(p)}g_0\lambda_1(p^2,0,p^2)\gamma_\mu \, .
\label{eq:vtx-asym-corr}
\end{equation}

As previously mentioned, in the Landau gauge we calculate the
transverse-projected vertex (\ref{eq:vtx-amp-trans}).  In the
asymmetric case, this becomes
\begin{align}
P_{\mu\nu}(q)\Lambda_\nu(p,q=0)
 & = \delta_{\mu\nu}\Lambda_\nu^{(ST)}(p,0) 
 + \Lambda_\mu^{(T)}(p,0) 
 = \Lambda^{(ST)}_\mu(p,0) \, ,
\end{align}
since $\Lambda_\mu^{(T)}(p,0)=0$ and $P_{\mu\nu}(0)=\delta_{\mu\nu}$ as
discussed on p.~\pageref{proj0}.
The lattice, finite-volume version of this is
\begin{align}
T_{\mu\nu}(q)\Lambda_\nu(p,q=0)\bigm|_{p_\mu=0}
 &= T_{\mu\mu}(0)\Lambda^{(ST)}_\mu(p,0)\bigm|_{p_\mu=0} 
 = -i\frac{T_{\mu\mu}(0)}{\zz(p)}g_0\lambda_1(p^2,0,p^2)\gamma_\mu \, .
\label{eq:TLlat1asym}
\end{align}

\subsection{Symmetric kinematics}
\label{sec:tree-sym}

The second case we consider is the (symmetric) case where $2p+q=0$,
i.e.\ $p=-k$ or equivalently $q=-2p$.
In this limit, the tree-level vertex (\ref{eq:vtx-tree}) becomes (for
ease of notation we will here set the lattice spacing $a=1$)
\begin{equation}
\Lambda^{a(0)}_{0,\mu}(p,-2p) \equiv -ig_0\Gamma^{a(0)}_{0,\mu}(p,-2p) 
 = -ig_0t^a\left(\gm + i\csw\sum_\nu\smn C_\mu(p)\Kt_\nu(p)\right) \, ,
\end{equation}
and repeating the same procedure as in the asymmetric case, the
improved vertex (\ref{eq:tree-imp}) takes the form
\begin{equation}
\Lambda_{I,\mu}^{a(0)}(p,-2p) = \bmp S_I^\z(p)^{-1}S_0^\z(p)
 \Lambda_{0,\mu}^{a(0)}(p,-2p)S_0^\z(-p)S_I^\z(-p)^{-1} \, .
\end{equation}
This gives us
\begin{multline}
\frac{D_I^2}{1+\bm}\Gamma^\z_{I,\mu}(p,-2p) =
\Bigl[B_V^2-A_V^2K^2+2A_VB_V\csw(\KdK)C_\mu\Bigr]\gamma_\mu
 + 2A_V^2\Kslash K_\mu\\
 - 2A_VB_V\csw\Ktslash\Kt_\mu - 2iA_VB_V\sum_\nu\smn K_\nu +
i\csw(A_V^2K^2+B_V^2)C_\mu\sum_\nu\smn\Kt_\nu \\
- 2i\csw A_V^2(\KdK)\Kt_\mu
 + 2i\csw A_V^2(\KdK)C_\mu\Kslash\gamma_\mu
 - 2i\csw A_V^2\Kt_\mu\sum_{\nu\lambda}\snl K_\nu\Kt_\lambda \, ,
\label{eq:sym-expand}
\end{multline}
where we have written $K=K(p),\Kt=\Kt(p),C=C(p)$ and
\begin{align}
A_V \equiv A_V(p) &= \bmp\bigl(m+\half Q^2(p)\bigr) - B_I(p) \, , \\
B_V \equiv B_V(p) &= \bmp K^2(p)
 + \bigl(m+\half Q^2(p)\bigr)B_I(p) \, .
\end{align}
If we concentrate on the part of this that becomes proportional to
$\lambda_1$ and $\tau_3$ in the continuum, the
lattice expression may be decomposed as
\begin{equation} 
\Gamma^\z_{I,\mu}(p,-2p) = \lambda_1^\z\gamma_\mu
 - 4\tau_3^\z(K^2\gamma_\mu-\Kslash K_\mu)
 - 4\tilde{\tau}_3^\z(\KdK C_\mu\gamma_\mu-\Ktslash\Kt_\mu) +
\ldots \label{eq:sym-decompose}
\end{equation}
We can read off the tree-level form factors from (\ref{eq:sym-expand}),
\begin{align}
\lambda_1^\z(p^2,4p^2,p^2) & = \phantom{-}
 \frac{1+\bm}{D_I(p)^2}\Bigl(A_V^2(p)K^2(p)+B_V^2(p)\Bigr)
 = 1/\zz(p) \, ; \label{eq:tree-imp-sym} \\
\tau_3^\z(p^2,4p^2,p^2) & =
 \phantom{-}\frac{1+\bm}{2D_I^2(p)}A_V^2(p) \, ; \\
\tilde{\tau}_3^\z(p^2,4p^2,p^2) & =
 -\frac{1+\bm}{2D_I^2(p)}\csw A_V(p)B_V(p) \, .
\end{align}
As in the asymmetric case,  we compute the transverse-projected vertex
(\ref{eq:vtx-amp-trans}) in Landau gauge.  With the decomposition
(\ref{eq:sym-decompose}), it reads (for sufficiently large $q$)
\begin{align}
\Gamma^P_\mu(p,-2p) & \equiv P^{\text{lat}}_{\mu\nu}(q)\Gamma_{\nu}(p,q=-2p) =
\Biggl(\delta_{\mu\nu}-\frac{Q_\mu(2p)Q_\nu(2p)}{Q^2(2p)}\Biggr)\Gamma_\nu(p,-2p)
 \notag \\ 
& = \Biggl(\delta_{\mu\nu}-\frac{K_\mu(p)K_\nu(p)}{K^2(p)}\Biggr)\Gamma_\nu(p,-2p) 
 \notag \\
 & = \Bigl(\lambda_1/K^2 - 4\tau_3\Bigl)(K^2\gamma_\mu-\Kslash K_\mu) 
 - 4\tilde{\tau}_3(\KdK C_\mu\gamma_\mu - \Ktslash\Kt_\mu) + \ldots
\label{eq:lat-sym-trans}
\end{align}
This is the lattice equivalent of the projection (\ref{eq:project}).
From this we obtain the transverse-projected, lattice equivalent of
(\ref{eq:sym-k1}),
\begin{equation}
\begin{split}
h_1(4p^2) = &\; -\quarter\sum_\mu\Im\Tr\gm\Lambda^T_\mu(p,-2p)\\
= &\; 3g_0\Bigl(\lambda_1(p^2,4p^2,p^2)-4K^2\tau_3(p^2,4p^2,p^2)\Bigr) \\
 & - 4g_0\Bigl(4\KdK-\Kt^2-\half Q^2\KdK\Bigr)\tilde{\tau}_3(p^2,4p^2,p^2)
 \equiv 3g_0\lambda_1^{\prime\text{lat}} \, .
\end{split}
\end{equation}
The tree-level corrected $\lambda'_1$ is therefore
\begin{equation}
\begin{split} 
\lambda'_1(p^2,4p^2,p^2)
 & = \frac{\lambda_1^{\prime\,\text{lat}}(p^2,4p^2,p^2)}
               {\lambda_1^{\prime\z}(p^2,4p^2,p^2)} \\
 & = \frac{h_1(4p^2)/g_0}
 {3(\lambda_1^\z-4K^2\tau_3^\z)
 - 4(4\KdK-\Kt^2-\half Q^2\KdK)\tilde{\tau}_3^\z} \, ,
\end{split}
\end{equation}
where the $p$-dependence in the denominator on the last line is
implicit.

\section{One-loop expressions}
\label{sec:one-loop}

In this section all the expressions will be given in Minkowski space.
In the \MSB\ scheme, the one-loop contributions $\Sigma_1^\ol,
\Pi^\ol$ and $\lambda_1^\ol$  to the quark and gluon
self-energy and the vertex component $\lambda_1$ in the \MOMT\
kinematics are given by \cite{Braaten:1981dv,Davydychev:2000rt}
\begin{multline}
\Sigma_1^\ol(p^2;\mu) = \frac{\gms^2(\mu)}{16\pi^2} C_F\Bigg\{ 
\xi\left[ 1+\frac{m^2}{p^2}-\ln\frac{m^2-p^2}{\mu^2}
         +\frac{m^4}{p^4}\ln\left(1-\frac{p^2}{m^2}\right) \right]\\
+\frac{m^2}{p^2}\left(1 -\frac{m^2}{p^2}\right)\ln\left(1-\frac{p^2}{m^2}\right)
\Bigg\} \,;
\label{eq:zquark-1loop}
\end{multline}
\vspace{-2mm}
\begin{multline}
\Pi^\ol(p^2;\mu) = \frac{\gms^2(\mu)}{16\pi^2}\Biggl\{
 \biggl[-\frac{97}{36}-\half\xi-\quarter\xi^2 +
 \Bigl(\frac{13}{6}-\frac{\xi}{2}\Bigr)\ln\frac{-p^2}{\mu^2}\biggr]C_A \\
 + \frac{4}{3}T_R N_f\biggl[
 \frac{1}{3}\,\left(5+12\frac{m^2}{p^2}\right)-\ln\frac{m^2}{\mu^2}\\ 
\phantom{\frac{4}{3}T_R N_f\biggl[}
 +\left(1+\frac{2m^2}{p^2}\right)\left(1-\frac{4m^2}{p^2}\right)^{1/2}
 \ln\frac{\left(1-\frac{4m^2}{p^2}\right)^{1/2}-1}
         {\left(1-\frac{4m^2}{p^2}\right)^{1/2}+1}\biggr]\Biggr\} \,;
\label{eq:zgluon-1loop}
\end{multline}
\vspace{-2mm}
\begin{multline}
\lambda^\ol_1(p^2,0,p^2;\mu)=\frac{\gms^2(\mu)}{16 \pi^2}
\biggl\{\xi C_F \left(1+\frac{m^2}{p^2}\right)
 + \frac{C_A}{4}\left[(3+\xi)+(1-\xi)\frac{m^2}{p^2}\right] \\
 - \left[\xi C_F + (3+\xi)\frac{C_A}{4}\right]\ln\frac{m^2-p^2}{\mu^2} \\
 + \left[\xi C_F + (1-\xi)\frac{C_A}{4}\right]
  \frac{m^4}{p^4}\ln\left(1-\frac{p^2}{m^2}\right) \biggr\} \,.
\label{eq:lambda1-1loop}
\end{multline}
Setting $p^2=-\mu^2$, this gives the following expression for the
\MOMT\ running coupling $\gmomt(\mu)$, in Landau gauge ($\xi=0$) for
$N_f=0$,
\begin{multline}
\gmomt(\mu) = \gms(\mu)\left[1+\lambda_1^\ol(-\mu^2,0,-\mu^2;\mu)
 -\Sigma_1^\ol(-\mu^2;\mu)-\half\Pi^\ol(-\mu^2;\mu)
+ \order(g^4)\right] \\
 = 
\gms(\mu)\Biggl[1+\biggl(\frac{151}{24}-\frac{3}{4}\frac{m^2}{\mu^2}
 - \frac{9}{4}\ln\Bigl(1+\frac{m^2}{\mu^2}\Bigr) \\
 + \frac{m^2}{\mu^2}\ln\Bigl(1+\frac{\mu^2}{m^2}\Bigr)
    \Bigl[\frac{4}{3} + \frac{25}{12}\frac{m^2}{\mu^2}\Bigr]\biggr)
 \frac{\gms^2(\mu)}{16\pi^2} + \order(g^4)\Biggr] \, .
\label{eq:gmomt-ms-app}
\end{multline}

In the \MOMB\ kinematics, the one-loop contributions to
$\lambda_1$ and $\tau_3$ are given by
\begin{multline}
\lambda_1^\ol(s^2,4s^2,s^2;\mu)=\\
\frac{\gms^2(\mu)}{16\pi^2}\Biggl\{
 \frac{5}{4}C_A + (C_F+\frac{3}{4}C_A)\xi
 + \left[C_F\xi+\frac{C_A}{4}(1-\xi)\right]\frac{m^2}{s^2}
 - \biggl[(C_F-\frac{C_A}{2})\xi +
 \frac{C_A}{4}(1+\xi)\frac{s^2+4m^2s^2+m^4}{s^2(s^2+m^2)} \biggr]
 \ln\frac{m^2-s^2}{\mu^2} \\
 - \frac{C_A}{2}(1+\xi)\frac{s^2}{s^2+m^2}\ln\frac{-4s^2}{\mu^2}
  + \frac{C_A}{4}(1+\xi)\frac{m^2}{s^2}\ln\frac{m^2}{\mu^2} \\
 + \biggl[ (C_F-\frac{C_A}{2})\xi\frac{m^2}{s^2}
   + \frac{C_A}{4}(1+\xi)\frac{s^2+4m^2s^2+m^4}{s^2(s^2+m^2)} \biggr]
  \frac{m^2}{s^2}\ln\Bigl(1-\frac{s^2}{m^2}\Bigr)
 \Biggr\} \,;
\end{multline}
\begin{multline}
\tau_3^\ol(s^2,4s^2,s^2;\mu) = \\
\frac{\gms^2(\mu)}{16\pi^2}\frac{1}{12s^2}\Biggl\{
 -(2C_F-\frac{5}{2}C_A)(2-\xi) - \frac{C_A}{2}\xi^2
 - \Bigl[4C_F + C_A(1-\xi)\Bigr]\frac{m^2}{s^2} \\
 - (2C_F-C_A)\Bigl[2+2\xi+\frac{m^2(5+\xi)}{s^2-m^2}\Bigr]
 \left[\sqrt{1-\frac{m^2}{s^2}}
  \ln\frac{\sqrt{1-\frac{m^2}{s^2}}+1}{\sqrt{1-\frac{m^2}{s^2}}-1}
     +\ln\frac{m^2}{\mu^2}\right]\\
 + \biggl\{4C_F(1+\xi)-\frac{C_A}{2}(7-2\xi-\xi^2)
   + (4C_F-C_A)\Bigl[\frac{m^2}{s^2}\xi+\frac{3m^2}{s^2-m^2}\Bigr] \\
   - C_A\frac{m^2}{s^2+m^2}
     \Bigl[\frac{s^2}{s^2+m^2}(5-4\xi-\xi^2)+\frac{m^2}{s^2}(1+\xi)\Bigr]
 \biggr\}\times \\
 \times \left[
  \ln\frac{m^2-s^2}{\mu^2}
   - \frac{m^2}{s^2}\ln\left(1-\frac{s^2}{m^2}\right)
\right]
 +\frac{C_A}{2}\frac{s^2}{(s^2+m^2)^2}
 \left[s^2(3-6\xi-\xi^2)+m^2(13-14\xi-3\xi^2)\right]\,
                          \ln\frac{-4s^2}{\mu^2} \Biggr\} \,.
\end{multline}
In Landau gauge, for SU(3), we find in the massless limit that
\begin{equation}
\begin{split}
\lambda_1^{\prime\ol}(s^2,4s^2,s^2;\mu) = &
 \lambda_1^\ol(s^2,4s^2,s^2;\mu) + 4s^2\tau_3^\ol(s^2,4s^2,s^2;\mu) \\
 = & \frac{\gms^2(\mu)}{16\pi^2}\left(\frac{251}{36} 
  + \frac{4}{9}\ln2 - \frac{9}{4}\ln\frac{-s^2}{\mu^2}\right) \, .
\end{split}
\end{equation}
Setting $s^2=-\mu^2$, analogously to the \MOMT\ scheme, we
find the \MOMB\ running coupling at asymptotically large momenta to be
\begin{equation}
\begin{split}
\gmomb(\mu) & = \gms(\mu)\left[1 + \lambda_1^{\prime\ol}
 -\Sigma_1^\ol - \half\Pi^\ol + \order(g^4)\right] \\
& = \gms(\mu)\Biggl\{
1+\bigg(\frac{4}{9}\ln2+\frac{793}{72}\bigg)\,
                 \frac{\gms(\mu)}{16\pi^2}+\order(g^4)
\Bigg\} \, .
\end{split}
\end{equation}


\providecommand{\href}[2]{#2}\begingroup\raggedright\endgroup

\end{document}